\documentclass[aps,pre,twocolumn,nofootinbib,tightenlines,superscriptaddress,nopacs,amsmath,amssymb,final,letterpaper]{revtex4-1}

\usepackage[utf8]{inputenc}
\usepackage{calc}
\usepackage{graphicx}
\usepackage{amsmath,amssymb,amsthm}
\usepackage{bbold}
\usepackage{bm}
\usepackage{color}
\usepackage[dvipsnames]{xcolor}
\usepackage{enumitem}
\usepackage{tikz}
\usepackage{float}

\newcommand{\abs}[1]{\left|#1\right|}

\newcommand{\MDL}{\mathrm{MDL}}
\newcommand{\cont}{\mathrm{cont}}
\newcommand{\comp}{\mathrm{comp}}

\usepackage[colorlinks=true,linkcolor=blue,urlcolor=blue,citecolor=blue,anchorcolor=blue]{hyperref}

\begin{document}
\title{Compressing network populations with modal networks reveals structural diversity}

\author{Alec \surname{Kirkley}}
\email{alec.w.kirkley@gmail.com}
\affiliation{Institute of Data Science, University of Hong Kong, Hong Kong}
\affiliation{Department of Urban Planning and Design, University of Hong Kong, Hong Kong}
\affiliation{Urban Systems Institute, University of Hong Kong, Hong Kong}

\author{Alexis \surname{Rojas}}
\affiliation{Department of Computer Science, University of Helsinki, Helsinki 00100, Finland}

\author{Martin \surname{Rosvall}}
\affiliation{Integrated Science Lab, Department of Physics, Umea University, SE-901 87 Umea, Sweden}

\author{Jean-Gabriel \surname{Young}}
\affiliation{Department of Mathematics and Statistics, University of Vermont, Burlington VT, USA}
\affiliation{Vermont Complex Systems Center, University of Vermont, Burlington VT, USA}

\date{\today}

\begin{abstract}
Analyzing relational data consisting of multiple samples or layers involves critical challenges: How many networks are required to capture the variety of structures in the data? And what are the structures of these representative networks?
We describe efficient nonparametric methods derived from the minimum description length principle to construct the network representations automatically. 
The methods input a population of networks or a multilayer network measured on a fixed set of nodes and output a small set of representative networks together with an assignment of each network sample or layer to one of the representative networks.
We identify the representative networks and assign network samples to them with an efficient Monte Carlo scheme that minimizes our description length objective.
For temporally ordered networks, we use a polynomial time dynamic programming approach that restricts the clusters of network layers to be temporally contiguous.
These methods recover planted heterogeneity in synthetic network populations and identify essential structural heterogeneities in global trade and fossil record networks.
Our methods are principled, scalable, parameter-free, and accommodate a wide range of data, providing a unified lens for exploratory analyses and preprocessing large sets of network samples. 

\end{abstract}

%%%%%%%%%%%%%%%%%%%%y%%%%%%%%%%%%%%%%%%%%%%%%%
%%%%%%%%%%%%%%%%%%%%%%%%%%%%%%%%%%%%%%%%%%%%%

\maketitle

\section{Introduction}
A common way to measure a network is to gather multiple observations of the connectivity of the same nodes.
Examples include the mobility patterns of a particular group of students encoded as a longitudinal set of co-location networks~\cite{eagle2009inferring,lehmann2019fundamental}, measurements of connectivity among the same brain regions for different individuals~\cite{sporns2010networks}, or the observation of protein-protein relationships through a variety of different interaction mechanisms \cite{stark2006biogrid}.
These measurements can be viewed as a multilayer network \cite{Kivela14} consisting of one layer for each measurement of all links between the nodes. For generality, we consider them as a \emph{population of networks}---a set of independent network measurements on the same set of nodes, either over time or across systems with consistent, aligned node labels. There often are regularities among such collections of measurements, but each sample may differ substantially from the next. 
Summarizing these measurements with robust statistical analyses can separate regularities from noise and simplify downstream analyses such as network visualization or regression ~\cite{butts2003network,Newman18b,young2020bayesian,Peixoto18,priebe2015statistical,arroyo2021inference,tang2018connectome,lunagomez2020modeling,wang2019common,young2021reconstruction}.

Most statistical methods for summarizing populations of networks share a similar approach.
They model all the members of a population as realizations of a single representative network~\cite{banks1994metric,butts2003network,Newman18a,Peixoto18,le2018estimating,lunagomez2020modeling}, which can be retrieved by fitting the model in question to the observed population. However, the strong assumption that a single ``modal'' network best explains the observed populations can lead to a poor representation of the data at hand~\cite{la2016gibbs,YKN22Clustering}.
For instance, accurately modeling a population of networks recording face-to-face interactions between elementary school pupils requires at least two representative networks if the data include networks observed during class and recess~\cite{stehle2011high}.
Modeling the measurements with a single network will most likely neglect essential variations in the pupil's face-to-face interactions, leading to similar oversights from summarizing a multimodal probability distribution with only its mean.

Recent research has examined related problems and led to, for example, methods for detecting abrupt regime changes in temporal series of networks~\cite{peel2015detecting,peixoto2018change}, pooling information across subsets of layers of multiplex networks~\cite{de2015structural}, and embedding nodes in common subspaces across network layers \cite{nielsen2018multiple,wang2019joint,arroyo2021inference}. Several recent contributions have addressed the problem of summarizing populations of networks when multiple distinct underlying network representations are needed, using mixtures of parametric models~\cite{stanley2016clustering,signorelli2020model,mantziou2021,yin2022finite,YKN22Clustering}, latent space models~\cite{durante2017nonparametric}, or generative models based on ad hoc graph distance measures~\cite{la2016gibbs}. 

These methods cluster network populations with good performance but have some significant drawbacks. None of the methods discussed, except Ref.~\cite{la2016gibbs}, outputs a single sparse representative network for each cluster but requires handling ensembles of network structures, making downstream applications such as network visualization or regression cumbersome. Most of these methods also require potentially unrealistic modeling assumptions about the structure of the clusters. For example, that stochastic block models or random dot product graphs can model all network structures in the clusters. Specifying a generative model for the modal structures also has the downside of often requiring complex and time-consuming methods to perform the within-cluster estimation. Perhaps most critically, existing approaches require either specifying the number of modes ahead of time or resorting to regularization with ad hoc penalties~\cite{mantziou2021,YKN22Clustering,de2015structural,durante2017nonparametric} not motivated directly by the clustering objective or approximative information  criteria~\cite{la2016gibbs,signorelli2020model,yin2022finite} poorly adapted to network problems. Overall, current approaches for clustering network populations do not provide a principled solution for model selection and often demand extensive tuning and significant computational overhead from fitting the model to several choices of the number of clusters.

Here we introduce nonparametric inference methods which overcome these obstacles and provide a coherent framework through which to approach the problem of clustering network populations or multiplex network layers, while extracting a representative modal network to summarize each cluster.
Our solution employs the minimum description length principle, which allows us to derive an objective function that favors parsimonious representations in an information-theoretic sense and selects the number and composition of representative modal networks automatically from first principles.
We first develop a fast Monte Carlo scheme for identifying the configuration of measurement clusters and modal networks that minimizes our description length objective.
We then extend our framework to account for special cases of interest: bipartite/directed networks and contiguous clusters containing all ordered networks from the earliest to the latest.
We show how to solve the latter problem in polynomial run time with a dynamic program~\cite{PAY2022}.
We demonstrate our methods in applications involving synthetic and real network data, finding that they can effectively recover planted network modes and clusters even with considerable noise. Our methods also provide a concise and meaningful summary of real network populations from applications in global trade and macroevolutionary research.

%%%%%%%%%%%%%%%%%%%%%%%%%%%%%%%%%%%%%%%%%%%%%
%%%%%%%%%%%%%%%%%%%%%%%%%%%%%%%%%%%%%%%%%%%%%

\section{Methods}

%%%%%%%%%%%%%%%%%%%%%%%%%%%%%%%%%%%%%%%%%%%%%
%%%%%%%%%%%%%%%%%%%%%%%%%%%%%%%%%%%%%%%%%%%%%

\subsection{Minimum description length objective}

For our clustering method, we rely on the minimum description length (MDL) principle: the best model among a set of candidate models is the one that permits the greatest compression---or shortest description---of a dataset~\cite{rissanen1978}.
The MDL principle provides a principled criterion for statistical model selection and has consequently been employed in various applications ranging from regression to time series analysis to clustering~\cite{hansen2001mdl}. A large body of research uses the MDL principle for clustering data, including studies on MDL-based methods for mixture models that accommodate continuous~\cite{georgieva2011cluster,tabor2014cross} and categorical data~\cite{li2004entropy}, as well as methods that are based on more general probabilistic generative models~\cite{narasimhan2005q}. The MDL approach has also been applied to complex network data, most notably for community detection algorithms to cluster nodes within a network~\cite{Rosvall07,Peixoto14a,kirkley2022spatial} and for decomposing graphs into subgraphs~\cite{koutra2014vog,wegner2014subgraph,bloem2020large,young2021hypergraph,bouritsas2021partition}, but also for clustering entire partitions of networks~\cite{Kirkley22Reps}. Our methods are similar in spirit to the one presented in ref.~\cite{Kirkley22Reps} for identifying representative community divisions among a set of plausible network partitions. Both approaches involve transmitting first a set of representatives and then the dataset itself by describing how each partition or network differs from its corresponding representative.
However, the methods differ substantially in their details since they address fundamentally different questions. 

We consider an experiment in which the initial data are a population of networks consisting of $S$ undirected, unweighted networks $\mathcal{D}=\{\bm{D}^{(1)}, ..., \bm{D}^{(S)}\}$ on a set of $N$ labeled nodes. 
The networks record, for instance, the co-location patterns among kids in a class of $N$ students over $S$ class periods.

We aim to summarize these data with $K$ modal networks $\mathcal{A}=\{\bm{A}^{(1)},..., \bm{A}^{(K)}\}$ (also undirected and unweighted) on the same set of nodes, with associated clusters of networks $\mathcal{C}=\{C_1,...,C_K\}$, where $C_k$ comprises networks that are similar to the mode $\bm{A}^{(k)}$.
This summary would allow researchers to, for instance, perform all downstream network analyses on a small set of representative networks---the modes---instead of a large set of networks likely to include measurement errors and from which it is difficult to draw valid conclusions.

We assume for simplicity of presentation that all networks $\mathcal{D}$ and $\mathcal{A}$ have no self- or multi-edges, although we can account for them straightforwardly.
While $K$ can be fixed if desired, we assume that it is unknown and must be determined from regularities in the data.

To select among all the possible modes and assignments of networks to clusters, we employ information theory and construct an objective function that quantifies how much information is needed to communicate the structure of the network population $\mathcal{D}$ to a receiver.
Clustering networks in groups of mostly similar instances allows us to communicate the population $\mathcal{D}$  efficiently in three steps: first the modes, then the clusters, and finally the networks $\mathcal{D}$ themselves as a series of small modifications to the modes $\mathcal{A}$.
The MDL principle tells us that any compression achieved in this way reveals modes and clusters that are genuine regularities of the population rather than noise~\cite{rissanen1978}.

We first establish a baseline for the code length: the number of bits needed to communicate $\mathcal{D}$ without using any regularities.
One way to do this is to first communicate the parameters of the population at a negligible information cost (size $S$,  number of nodes $N$, and the total number of edges $E$ in all networks of $\mathcal{D}$) and then transmit the population $\mathcal{D}$ directly.
There are ${N\choose 2}$ possible edge positions in each of the $S$ undirected networks in $\mathcal{D}$, or $S{N\choose 2}$ possible edge positions for the whole population. So these networks can be configured in ${S{N\choose 2} \choose E}$ ways.
It thus takes approximately
\begin{align}
    \mathcal{L}_{0}(\mathcal{D}) = \log {S{N\choose 2} \choose E}
\end{align}
bits to transmit these networks to a receiver.
(We use the convention $\log\equiv \log_2$ for brevity.)
Applying Stirling's approximation $\log x!\approx x\log x - x/\ln(2)$, we obtain 
\begin{align}
    \label{eq:L0}
    \mathcal{L}_{0}(\mathcal{D}) \approx S{N\choose 2} H_b\left(\frac{E}{S{N\choose 2}}\right)
\end{align}
written in terms of the binary Shannon entropy
\begin{align}
    H_b(p) = -p\log p - (1-p)\log (1-p).     
\end{align}

\begin{figure}
    \centering
    \includegraphics[width=1\linewidth]{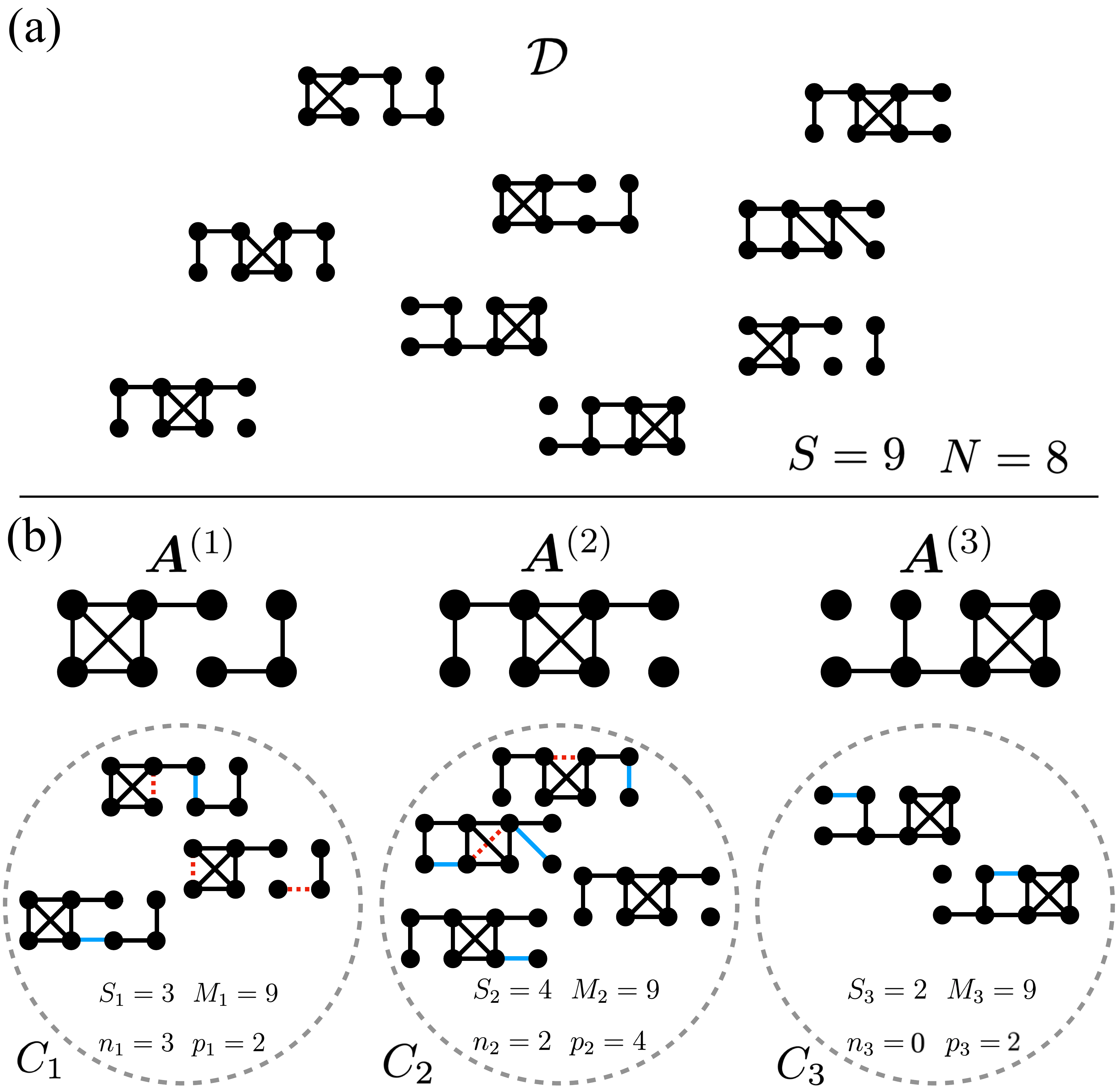}
    \caption{\textbf{Information transmission scheme}. \textbf{(a)} Example population of networks $\mathcal{D}$, with $S=9$ networks of $N=8$ nodes each. \textbf{(b)} Representative modes $\{\bm{A}^{(k)}\}$ with their corresponding clusters of networks $\{C_k\}$. First, each mode network is transmitted individually in its entirety, with information content $\mathcal{L}(\bm{A}^{(k)})$ given by Eq.~\eqref{eq:LAk}. Then, networks in the population are assigned to disjoint clusters surrounding each mode, requiring an information content given by Eq.~\eqref{eq:LC}. Finally, all the networks $\bm{D}^{(s)}$ in each cluster $C_k$ are transmitted, given the number of false-negative and false-positive edges $n_k$ and $p_k$ in the cluster (represented with dotted red and solid blue lines, respectively). The information content of this step is given by $\ell_k$ in Eq.~\eqref{eq:LDk}. Different choices of clusters and modes lead to a different total information content, and the aim is to identify the clusters and modes that minimize this information content.}
    \label{fig:diagram}
\end{figure}

In practice, we expect to need many fewer bits than $\mathcal{L}_0$ to communicate $\mathcal{D}$, because the population of networks will often have regularities.
We propose a multi-part encoding that identifies such regularities by grouping similar networks in clusters $\mathcal{C}$ with modes $\mathcal{A}$, which proceeds as follows.
First, we send a small number of modes $\mathcal{A}$ in their entirety, which ideally captures most of the heterogeneity in the population $\mathcal{D}$.
This step is costly but will save us information later.
We then send the network clusters $\mathcal{C}$ by transmitting the cluster label of each network $s\in \mathcal{D}$. Finally, we transmit the edges of networks in each cluster, using the already transmitted modes as a starting point to compress this part of the encoding significantly.
The expected code length can be quantified using simple combinatorial expressions, and the configuration of modes $\mathcal{A}$ and clusters $\mathcal{C}$ that minimizes the total expected code length---the MDL configuration---provides a succinct summary of the data $\mathcal{D}$. 
Figure~\ref{fig:diagram} summarizes the transmission process and the individual description length contributions.

The expected length of this multi-part encoding is the sum of the description length of each part of the code that has significant communication costs.
The modes are the first objects that incur such costs.
Following the same reasoning as before, we denote the number of edges in mode $k$ as $M_k$ and conclude that we can transmit the positions of the occupied edges in mode $\bm{A}^{(k)}$ using approximately
\begin{align}
    \label{eq:LAk}
    \mathcal{L}(\bm{A}^{(k)}) = \log {{N\choose 2}  \choose M_k} \approx {N\choose 2}H_b\left(\frac{M_k}{{N\choose 2}}\right)
\end{align}
bits, where the second expression results from a Stirling approximation as in Eq.~\eqref{eq:L0}.
We can therefore transmit all the modes with a total code length of
\begin{align}
    \label{eq:LA}
    \mathcal{L}(\mathcal{A})=\sum_{k=1}^{K}\mathcal{L}(\bm{A}^{(k)})
\end{align}
bits.

The next step is to transmit the cluster label $k$ of each network in $\mathcal{D}$.
For this part of the code, we first send the number of networks $S_k$ in each cluster $k=1,...,K$ at a negligible cost and then specify a particular clustering compatible with these constraints.
The multinomial coefficient ${S\choose S_1\ S_2\ \cdots\ S_k}$ gives the total number of possible combinations of these cluster labels. The information content of this step is thus
\begin{align}
    \label{eq:LC}
    \mathcal{L}(\mathcal{C}) = \log {S\choose S_1\ S_2\ \cdots\ S_k}
    \approx S \ H\Big(\{S_k/S\}\Big),
\end{align}
where we again use the Stirling approximation and where
\begin{align}
    H\Big(\{q_k\}\Big) = -\sum_{k=1}^{K}q_k\log q_k    
\end{align}
is the Shannon entropy of a distribution $\{q_k\}$.

Finally, we transmit the network population $\mathcal{D}$ by sending the \emph{differences} between the networks in each cluster and their associated mode.
To calculate the length of this part of the code, we focus on a particular cluster $C_k$ and count the number of times we will have to remove an edge from the mode $\bm{A}^{(k)}$ when specifying the structure of networks in its cluster using $\bm{A}^{(k)}$ as a reference.
We call these edges \emph{false negatives}, and count them as
\begin{align}
    \label{eq:nk}
    n_k = \sum_{s\in C_k} \abs{\bm{A}^{(k)} \setminus \bm{D}^{(s)}},
\end{align}
where we interpret $\bm{D}^{(s)}$ and $\bm{A}^{(k)}$ as sets of edges, so the summand is the number of edges in mode $k$ that are not in the network $s$.
Similarly, we also require the number of edges that occur in the networks of cluster $k$ but not in the mode---the number of \emph{false positives},
\begin{align}
    \label{eq:pk}
    p_k = \sum_{s\in C_k}\abs{\bm{D}^{(s)}\setminus \bm{A}^{(k)}}.
\end{align}
Like the cluster sizes $S_k$ and edge counts per cluster $M_k$, the pairs $(n_k,p_k)$ can be communicated to the receiver at a comparatively negligible cost, and we ignore them in our calculations.

To estimate the information content of this part of the transmission, we count the number of configurations of false-negative and false-positive edges in $C_k$.
Focusing first on the false negatives---the edges that must be deleted---we count that of the $S_kM_k$ edges in the $S_k$ copies of the mode of cluster $k$, $n_k$ will be false-negative edges that can be configured in ${S_kM_k \choose n_k}$ ways. 
Similarly, using the shorthand $ M_k^\ast ={N\choose 2} - M_k$ to denote the unoccupied pairs of nodes in the mode $k$, there are $S_k M_k^\ast$ locations in which we must place $p_k$ false-positive edges, for a total of ${S_k M_k^\ast  \choose p_k}$ possible configurations of false-positive edges.
The total information content required for transmitting the locations of the false-negative and false-positive edges of every network in cluster $k$ is thus
\begin{align}
    \ell_k &:= \mathcal{L}(\{\mathcal{D}^{(s)}|s\in C_k\}\vert \bm{A}^{(k)})\nonumber\\
    &=\log {S_kM_k \choose n_k} + \log {S_k M_k^\ast  \choose p_k},
    \label{eq:LDk}
\end{align}
which we approximate as
\begin{equation}
    \ell_k \approx S_kM_k H_b\left(\frac{n_k}{S_kM_k}\right) 
    +S_k M_k^\ast  H_b\left(\frac{p_k}{S_k M_k^\ast }\right).    
\end{equation}
Summing over all clusters,
\begin{align}
    \label{eq:LDgivenAC}
    \mathcal{L}(\mathcal{D}\vert \mathcal{A},\mathcal{C}) = \sum_{k=1}^{K}\ell_k,
\end{align}
we obtain the total information content of the final step in the transmission process.\\

We obtain the total description length $\mathcal{L}(\mathcal{D})$ by adding the contributions of Eqs.~\eqref{eq:LA}, \eqref{eq:LC}, and~\eqref{eq:LDgivenAC}, as
\begin{align}
\mathcal{L}(\mathcal{D}) = \mathcal{L}(\mathcal{A}) + \mathcal{L}(\mathcal{C}) + \mathcal{L}(\mathcal{D}\vert \mathcal{A},\mathcal{C}). 
\end{align}
This objective function allows for efficient optimization because we can express it as a sum of the cluster-level description lengths
\begin{equation}
\label{eq:Lk}
\mathcal{L}_k(\bm{A}^{(k)},C_k)=\mathcal{L}(\mathcal{A}^{(k)}) + S\log \left(\frac{S}{S_k}\right)  + \ell_k,    
\end{equation}
giving
\begin{align}
\label{eq:DLtotal}
\mathcal{L}(\mathcal{D}) =  \sum_{k=1}^{K}\mathcal{L}_k(\bm{A}^{(k)},C_k).    
\end{align}
Equations~\eqref{eq:LAk} and~\eqref{eq:LDk} provide explicit expressions for $\mathcal{L}(\mathcal{A}^{(k)})$ and $\ell_k$.

Equation~\eqref{eq:DLtotal} gives the total description length of the data $\mathcal{D}$ under our multi-part transmission scheme. By minimizing this objective function we identify the best configurations of modes $\mathcal{A}$ and clusters $\mathcal{C}$. 
A good configuration $\{\mathcal{A},\mathcal{C}\}$ will allow us to transmit a large portion of the information in $\mathcal{D}$ through the modes alone.
If we use too many modes, the description length will increase as these are costly to communicate in full.
And if we use too few, the description length will also increase because we will have to send lengthy messages describing how mismatched networks and modes differ.
Hence, through the principle of parsimony, Eq.~\eqref{eq:DLtotal} favors descriptions with the number of clusters $K$ as small as possible but not smaller.

This framework can be modified to accommodate populations of bipartite or directed networks. For the bipartite case, we make the transformations ${N\choose 2}\to N_1N_2$ and $ M_k^\ast \to N_1N_2-M_k$, where $N_1$ and $N_2$ are the numbers of nodes in each of the two groups. This modification reduces the number of available positions for potential edges. Similarly, for the directed case, we can make the transformations ${N\choose 2}\to N(N-1)$ and $ M_k^\ast \to N(N-1)-M_k$, which increases the number of available edge positions.

%%%%%%%%%%%%%%%%%%%%%%%%%%%%%%%%%%%%%%%%%%%%%
%%%%%%%%%%%%%%%%%%%%%%%%%%%%%%%%%%%%%%%%%%%%%

\subsection{Optimization and model selection}
\label{sec:algorithm}
Since Eq.~\eqref{eq:DLtotal} has large support, is not convex, and has many local optima, a stochastic optimization method is a natural choice for finding reasonable solutions rapidly.
We exploit the objective function's decoupling into a sum over clusters $k$ and implement an efficient merge-split Monte Carlo method for the search \cite{peixoto2020merge,Kirkley22Reps}.
The method greedily optimizes $\mathcal{L}(\mathcal{D})$ using moves that involve merging and splitting clusters of networks $D^{(s)}\in\mathcal{D}$.

Our merge-split algorithm minimizes the description length in Eq.~\eqref{eq:DLtotal} by performing one of the following moves selected uniformly at random and accepting the move as long as it results in a reduction of the description length~\eqref{eq:DLtotal}: 
\begin{enumerate}
    \item \textbf{Reassignment}: Pick a network $s$ at random and move it from its current cluster $C_k$ to the cluster $C_{k'}$ that results in the greatest decrease in the description length. Compute the modes $\bm{A}^{(k)}$ and $\bm{A}^{(k')}$ that minimize the cluster-level description lengths $\mathcal{L}_k(\bm{A}^{(k)},C_k)$ and $\mathcal{L}_{k'}(\bm{A}^{(k')},C_{k'})$ using Eq.~\eqref{eq:Lk} and the procedure described below, conditioned on the networks in $C_k$ and $C_{k'}$. 
    \item \textbf{Merge}: Pick two clusters $C_{k'}$ and $C_{k''}$ at random and merge them into a single cluster $C_k$. 
    Compute the mode $\bm{A}^{(k)}$ that minimizes the cluster-level description length $\mathcal{L}_k(\bm{A}^{(k)},C_k)$ using Eq.~\eqref{eq:Lk} and the procedure described below, conditioned on the networks in $C_k$. Finally, compute the change in the description length that results from this merge.
    \item \textbf{Split}: Pick a cluster $C_{k}$ at random and split it into two clusters $C_{k'}$ and $C_{k''}$ using the following $2$-means algorithm. First assign every network in $C_k$ to the cluster $C_{k'}$ or $C_{k''}$ at random. Refine the assignments by successively moving every network to the cluster $C_{k'}$ or $C_{k''}$ that results in a greater decrease in the description length and compute
    the modes $\bm{A}^{(k')}$ and $\bm{A}^{(k'')}$ 
    that minimize the cluster-level description lengths $\mathcal{L}_{k'}(\bm{A}^{(k')},C_{k'})$ and $\mathcal{L}_{k''}(\bm{A}^{(k'')},C_{k''})$, conditioned on the networks now in $C_{k'}$ and $C_{k'}$. 
    After convergence of the $2$-means style algorithm, compute the change in the description length that results from this split of cluster $C_k$.
    \item \textbf{Merge-split}: Pick two clusters at random, merge them as in move 2, then perform move 3 on this merged cluster. These two moves in direct succession help reassign multiple networks simultaneously; their addition to the move set improves the algorithm's performance.  
\end{enumerate}

Since these moves modify only one or two clusters, the change in the global description length $\mathcal{L}(\mathcal{D})$ can be recomputed quickly as updates to the cluster-level description lengths in Eq.~\eqref{eq:Lk}.
Every time a mode is needed for these calculations, we use the mode that minimizes the cluster-level description length $\mathcal{L}_k(\bm{A}^{(k)},C_k)$ in Eq.~\eqref{eq:Lk}.
To find this optimal mode efficiently, we start with the ``complete'' mode
\begin{align}
    \label{eq:Acomp}
    \bm{A}_{\comp}^{(k)}=\bigcup_{s\in C_k}\bm{D}^{(s)},    
\end{align}
with an edge between nodes $i$ and $j$ if at least one network in the cluster contains the edge.
We then greedily remove edges from $\bm{A}_{\comp}^{(k)}$ in increasing order of occurrence in the networks of $C_k$---starting first with edges only found in a single network and going up from there---and update the cluster-level description length as we go.
After removing all edges from $\bm{A}_{\comp}^{(k)}$, the mode giving the lowest cluster-level description length is chosen as the mode for the cluster.
This approach is locally optimal under a few assumptions about the sparsity of the networks and the composition of edges in the clusters (see Appendix~\ref{sec:appendixmergesplit} for details).

We run the algorithm by starting with $K_0$ initial clusters (this choice has a negligible effect on the results, see Appendix~\ref{sec:tests}) and stop when a specified number of consecutive moves all result in rejections, indicating that the algorithm has likely converged.
The worst-case complexity of this algorithm is roughly $O(NS)$ (the worst case is a split move right at the start, see Appendix~\ref{sec:tests}).
Appendix~\ref{sec:appendixmergesplit} details the entire algorithm, and Appendix~\ref{sec:tests} provides additional tests of the algorithm, such as its robustness for different choices of $K_0$.

To diagnose the quality of a solution, we compute the \emph{inverse compression ratio} 
\begin{align}
\label{eq:ratio}
\eta(\mathcal{D}) = \mathcal{L}_{\MDL}(\mathcal{D})/\mathcal{L}_0(\mathcal{D}), 
\end{align}
where $\mathcal{L}_{\MDL}(\mathcal{D})$ is the minimum value of $\mathcal{L}(\mathcal{D})$ over all configurations of $\mathcal{A},\mathcal{C}$, given by the algorithm after termination, and $\mathcal{L}_0$ is given in Eq.~\eqref{eq:L0}. Equation~\eqref{eq:ratio} tells us how much better we can compress the network population $\mathcal{D}$ by using our multi-step encoding than by using the na\"ive fixed-length code to transmit all networks individually. If $\eta(\mathcal{D})<1$, our model compresses the data $\mathcal{D}$, and if $\eta(\mathcal{D})>1$, it does not because we waste too much information in the initial transmission steps.

%%%%%%%%%%%%%%%%%%%%%%%%%%%%%%%%%%%%%%%%%%%%%
%%%%%%%%%%%%%%%%%%%%%%%%%%%%%%%%%%%%%%%%%%%%%

\subsection{Contiguous clusters}
\label{sec:temporal}

In Sec.~\ref{sec:algorithm}, we described a merge-split Monte Carlo algorithm to identify the clusters $\mathcal{C}$ and modes $\mathcal{A}$ that minimize the description length in Eq.~\eqref{eq:DLtotal}. This algorithm samples the space of unconstrained partitions $\mathcal{C}$ of the network population $\mathcal{D}$. However, in many applications, particularly in longitudinal studies, we may only be interested in constructing contiguous clusters, where each cluster is now a set of  networks where adjacent indexes $s\in\{1,...,S\}$ indicate contiguity of some form (temporal, spatial, or otherwise). Such constraints reduce the space of possible clusterings $\mathcal{C}$ drastically, and we can minimize the description length exactly (up to the greedy heuristic for the mode construction) using a dynamical program~\cite{jackson2005algorithm,bellman2013dynamic,PAY2022}.

Before we introduce an optimization for this problem, we require a small modification to Eq.~\eqref{eq:Lk} for the cluster-level description length to accurately reflect the constrained space of ordered partitions $\mathcal{C}$ that we are considering.
In our derivation of the description length, we assumed that the receiver knows the sizes $\{S_k\}$ of the clusters in $\mathcal{C}$. If we transmit these sizes in the order of the clusters they describe, the receiver will also know the exact clusters $\mathcal{C}$, since knowing the sizes $\{S_k\}$ is equivalent to knowing the cluster boundaries in this contiguous case. We can therefore ignore the term $S\log (S/S_k)$ in Eq.~\eqref{eq:Lk} that tells us how much information is required to transmit the exact cluster configuration. This modification results in a new, shorter description length
\begin{align}
    \label{eq:Lkcont}
    \mathcal{L}^{(\cont)}_{k}(\bm{A}^{(k)},C_k)=\mathcal{L}(\mathcal{A}^{(k)}) + \ell_k
\end{align}
and a new global objective
\begin{align}
    \label{eq:Lcont}
    \mathcal{L}_{\cont}(\mathcal{D}) = \sum_{k}\mathcal{L}^{(\cont)}_{k}(\bm{A}^{(k)},C_k).
\end{align}
Since the objective in Eq.~\eqref{eq:Lcont} is a sum of independent cluster-level terms, minimizing this description length for contiguous clusters admits a dynamic programming algorithm solution~\cite{jackson2005algorithm,bellman2013dynamic,PAY2022} that can identify the true optima in polynomial time.

The algorithm is constructed by recursing on $\mathcal{L}^{(i)}_{\MDL}$, the minimum description length of the first $i$ networks in $\mathcal{D}$ according to Eq.~\eqref{eq:Lcont}.
Since the objective function decomposes as a sum over clusters, for any $j\in [1,S]$, the MDL can be calculated as
\begin{align}
    \label{eq:dynamic}
    \mathcal{L}^{(j)}_{\MDL} = \underset{i\in [1,j]}{\operatorname{min}} \left\{\mathcal{L}^{(i-1)}_{\MDL} + \mathcal{L}^{(\cont)}_k([i,j])  \right\},  
\end{align}
where we set the base case to $\mathcal{L}^{(0)}_{\MDL}=0$ and define $\mathcal{L}^{(\cont)}_k([i,j])$ as the description length of the cluster of networks with indices $\{i,...,j\}$, according to Eq.~\eqref{eq:Lkcont} with the mode computed with the greedy procedure described in Sec.~\ref{sec:algorithm}.
Once we recurse to $\mathcal{L}^{(S)}_{\MDL}$, we have found the MDL of our complete dataset, and keeping tab of the minimizing $i$ in Eq.~\eqref{eq:dynamic} for every $j$ allows us to reconstruct the clusters.

In practice, the recursion can be implemented from the bottom up, starting with $\mathcal{L}^{(1)}_{\MDL}$, then $\mathcal{L}^{(2)}_{\MDL}$, and so on.
The computational bottleneck for calculating $\mathcal{L}^{(j)}_{\MDL}$ is finding the modes of a cluster $j$ times for each evaluation of Eq.~\eqref{eq:dynamic} (once for each $i=1,...,j$), leading to an overall complexity $O(jN\log N)$ for this step. 
Summing over $j\in [1,S]$, the overall time complexity of the dynamic programming algorithm is $O(S^2N\log N)$, which we verify numerically in Appendix~\ref{sec:tests}.

%%%%%%%%%%%%%%%%%%%%%%%%%%%%%%%%%%%%%%%%%%%%%
%%%%%%%%%%%%%%%%%%%%%%%%%%%%%%%%%%%%%%%%%%%%%

\section{Results}

We test our methods on a range of real and synthetic example network populations. First, we show that our algorithms can recover synthetically generated clusters and modes with high accuracy despite considerable noise levels. Applied to worldwide networks of food imports and exports, we find a strong compression that uses the difference between categories of products and the locations in which they are produced. We then apply our method for contiguous clustering of ordered network populations to a set of networks representing the fossil record from ordered geological stages in the last 500 million years \cite{rojas_multiscale_2021}. We examine bipartite and unipartite representations of these systems and find close alignment between our inferred clusters and known global biotic transitions, including those triggered by mass extinction events. 

%%%%%%%%%%%%%%%%%%%%%%%%%%%%%%%%%%%%%%%%%%%%%
%%%%%%%%%%%%%%%%%%%%%%%%%%%%%%%%%%%%%%%%%%%%%

\subsection{Reconstruction of synthetic network populations}
\label{sec:synthetic}

To demonstrate that the methods presented in Secs.~\ref{sec:algorithm} and~\ref{sec:temporal} can effectively identify modes and clusters in network populations, we test their ability to recover the underlying modes and clusters generated from the heterogeneous population model introduced in ref.~\cite{YKN22Clustering}. We examine the robustness of these methods under varying noise levels that influence the similarity of the generated networks with the cluster's mode.   

The generative model in \cite{YKN22Clustering} supposes (using different notation) that we are given $K$ modes $\mathcal{A}$ as well as the cluster assignments $\mathcal{C}$ of the networks $\mathcal{D}$. Each network $s\in C_k$ is generated by first taking each edge $(i,j)\in \bm{A}^{(k)}$ independently and adding it to $\bm{D}^{(s)}$ with probability $\alpha_k$ (the \emph{true-positive rate}). Then, each of $ M_k^\ast $ possible edges absent from $\bm{A}^{(k)}$ is added to $\bm{D}^{(s)}$ with probability $\beta_k$ (the \emph{false-positive rate}). After performing this procedure for all clusters, the end result is a heterogeneous population of networks $\mathcal{D}$ with $K$ underlying modes, with noise in the networks $C_k$ surrounding each mode $\bm{A}^{(k)}$ determined by the rates $\alpha_k$ and $\beta_k$. The higher the true-positive rate $\alpha_k$ and the lower the false-positive rate $\beta_k$, the closer the networks in cluster $C_k$ resemble their corresponding mode $\bm{A}^{(k)}$. 

\begin{figure}
    \centering
    \includegraphics[width=0.95\columnwidth]{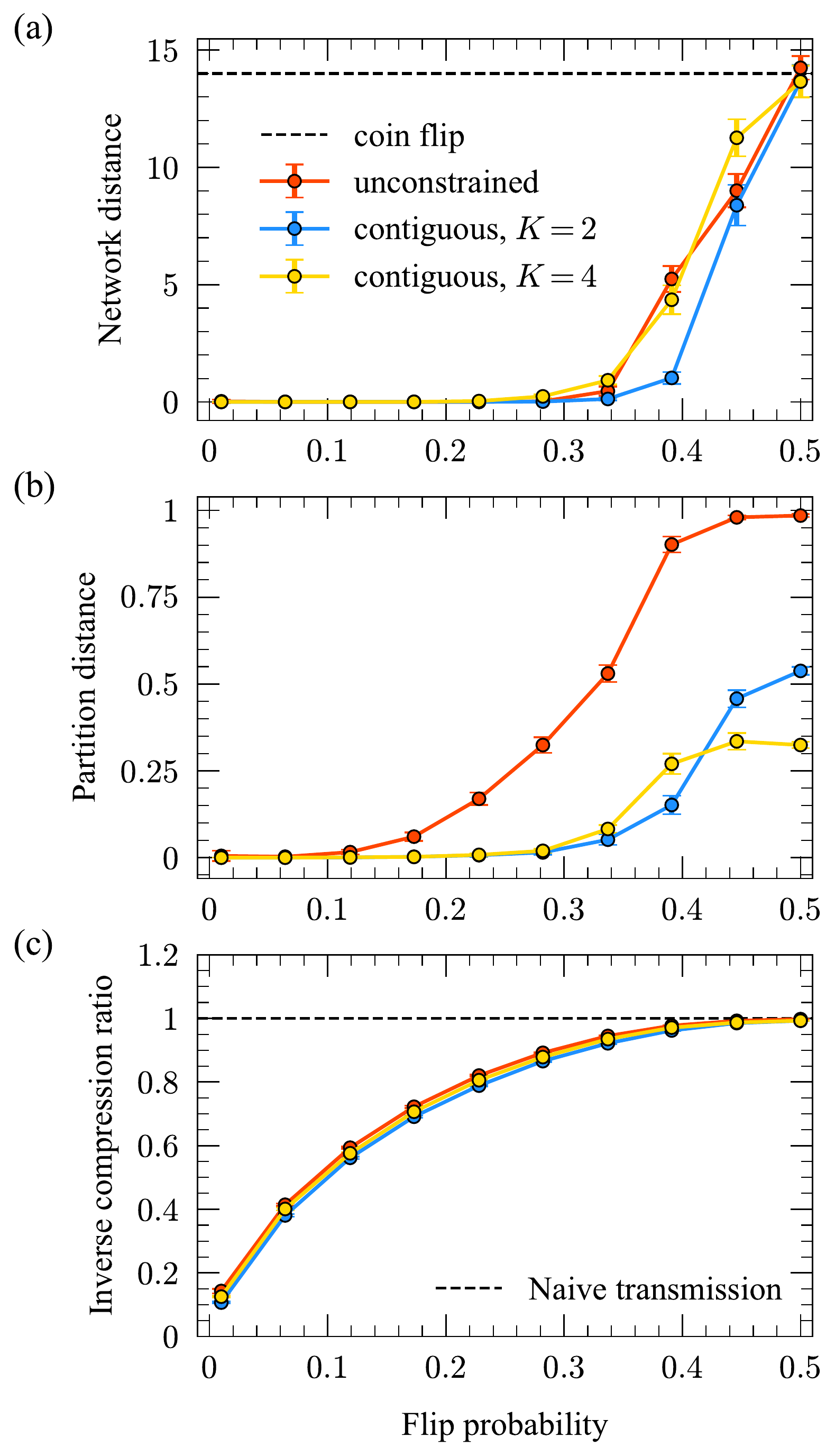}
    \caption{\textbf{Recovery of planted modes and their clusters in synthetic network populations}. Various aspects of the recovery performance are plotted for the three experiments described in Sec.~\ref{sec:synthetic}. \textbf{(a)} Network distance, as quantified by the average Hamming distance between the true and inferred and modes (modes 1 and 3 in Fig.~\ref{fig:diagram}), for various flip probabilities $p$. \textbf{(b)} Partition distance, given by the one minus the normalized mutual information between the true and inferred clusterings of the network population. \textbf{(c)} Inverse compression ratio, given in Eq.~\eqref{eq:ratio}. Each data point is an average over $200$ realizations of the population for the corresponding value of the flip probability, and error bars correspond to three standard errors in the mean. }
    \label{fig:recovery}
\end{figure}

Employing Bayesian inference of the modes and cluster assignments as in ref.~\cite{YKN22Clustering} involves adding prior probability distributions over the modes $\mathcal{A}$ and cluster assignments $\mathcal{C}$ to the heterogeneous network model \cite{YKN22Clustering}. With a specific choice of priors on the modes and cluster sizes, Eq.~\eqref{eq:DLtotal} is precisely the equation giving us the Maximum A Posteriori (MAP) estimators of $\mathcal{A}$ and $\mathcal{C}$ in this model. We defer the details of this correspondence to Appendix~\ref{sec:MAP}.

For our experiments, we use two modes, mode 1 and mode 3 from Fig.~\ref{fig:diagram}, as the planted modes $\mathcal{A}_{true}$ we aim to recover. To provide a single intuitive parameter quantifying the noise level in the generative model, we choose the true- and false-positive rates to satisfy $p=\beta_1=\beta_3=1-\alpha_1=1-\alpha_3$ for each run. Viewing the networks as binary adjacency matrices, the parameter $p$ corresponds to the probability of flipping entries of the matrix from 0 to 1 and vice-versa when constructing a network from its assigned cluster. We denote the parameter $p$ as the ``flip probability'' to emphasize this interpretation (same formulation as in ref.~\cite{YKN22Clustering}). A flip probability $p=0$ corresponds to clusters of networks identical to the cluster modes, and a flip probability of $p=0.5$ corresponds to completely random networks with no clustering in the population. We thus expect it to be easy to recover the planted modes $\mathcal{A}_{true}$ and clusters $\mathcal{C}_{true}$ for $p=0$, and the problem becomes more and more difficult as we approach $p=0.5$.   

We run three separate recovery experiments to test both the unconstrained and contiguous description length objectives in Eq.~\eqref{eq:DLtotal} and Eq~\eqref{eq:Lcont}, respectively. For the unconstrained objective, in each run we generate a population of $S$ networks from the model described above, with each network generated from either mode 1 or mode 3 at random with equal probability. We then identify the modes $\mathcal{A}_{\MDL}$ and clusters $\mathcal{C}_{\MDL}$ that minimize the objective in Eq.~\eqref{eq:DLtotal} using the merge-split algorithm detailed in Sec.~\ref{sec:algorithm} and Appendix~\ref{sec:appendixmergesplit}. For the recovery of contiguous clusters, in one experiment we generate $S/2$ consecutive networks from each mode so that the population consists of $K=2$ adjacent contiguous clusters. And in another experiment, we generate $S/4$ networks from mode 1, $S/4$ networks from mode 3, and repeat this so that there are $K=4$ adjacent contiguous clusters of the $S$ networks generated from the two distinct modes. For these two experiments, we run the dynamic programming algorithm detailed in Sec.~\ref{sec:temporal} to identify the modes $\mathcal{A}_{\MDL}$ and clusters $\mathcal{C}_{\MDL}$ that minimize the objective in Eq.~\eqref{eq:Lcont}. In all three experiments, we generate a population of $S=100$ networks, each constructed from its corresponding mode using the single flip probability $p$ to introduce true- and false-positive edges.

To quantify the mode recovery error, we use the network distance quantified by the average Hamming distance between the inferred modes $\mathcal{A}$ and the planted modes $\mathcal{A}_{true}$. As both of our algorithms automatically select the optimal number of clusters $K$, the number of modes we infer can differ from the true number ($K=2$ or $K=4$, depending on the experiment). In each experiment, we therefore choose the $K$ inferred modes in $\mathcal{A}$ with the largest corresponding clusters and compute the average Hamming distance between these and the true modes in $\mathcal{A}_{true}$. (Since there are $K!$ ways to choose the inferred mode labels, we choose the labeling that produces the smallest Hamming distance.) To measure the error between our inferred clusters $\mathcal{C}$ and the planted clusters $\mathcal{C}_{true}$ (the ``partition distance''), we use one minus the normalized mutual information \cite{vinh2010information}. We also compute the inverse compression ratio (Eq.~\eqref{eq:ratio}) to measure how well the network population can be compressed. We pick a range of values of $p$ to tune the noise level in the populations, and at each value of $p$ we average these three quantities over $200$ realizations of the model to smooth out noise due to randomness in the synthetic network populations. We choose $K_0=1$ for these experiments, but this choice has little to no effect on the results (see Appendix~\ref{sec:tests}).

\begin{figure}
    \centering
    \includegraphics[width=0.95\columnwidth]{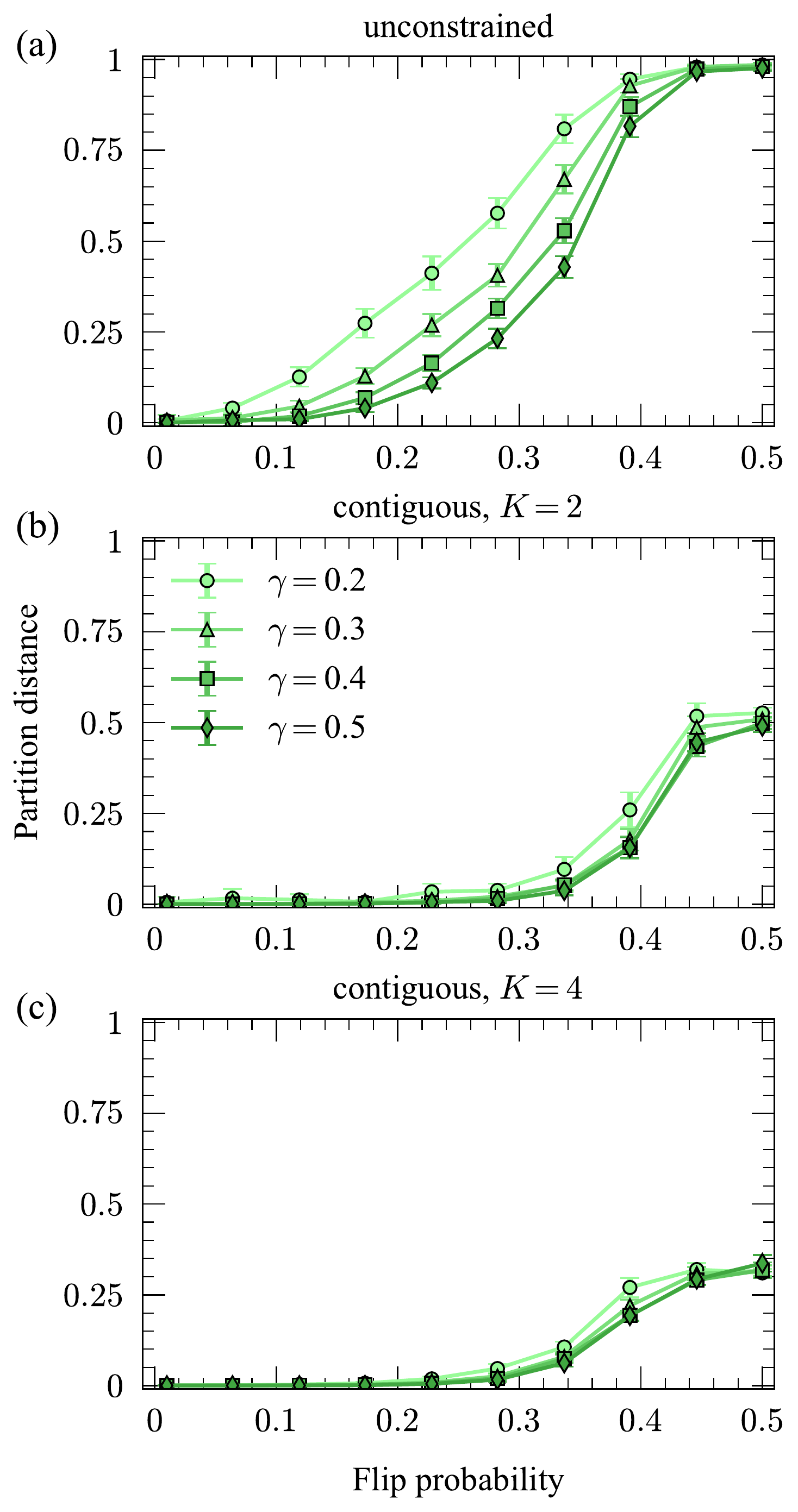}
    \caption{\textbf{Cluster recovery for different mode separations}. Partition distance between true and inferred clusters for \textbf{(a)} unconstrained clustering, \textbf{(b)} contiguous clustering with $K=2$, and \textbf{(c)} contiguous clustering with $K=4$ for various values of the mode separation $\gamma$. Each data point is an average over $200$ realizations of the population for the corresponding value of the flip probability, and error bars correspond to three standard errors in the mean. }
    \label{fig:recovery2}
\end{figure}

Figure~\ref{fig:recovery} shows the results of our first reconstruction experiment. 
The reconstruction performance gradually worsens as $p$ increases due to the increasing noise level in the sampled networks relative to their corresponding modes (Fig.~\ref{fig:recovery}a). In all experiments, the network distance reaches that expected for a completely random guess of the mode networks---a 50/50 coin flip to determine the existence of each edge, denoted by the dashed black line---when $p=0.5$. The results in Fig.~\ref{fig:recovery}a indicate that in both the unconstrained and contiguous cases, our algorithms are capable of recovering the modes underlying these synthetic network populations with high accuracy, even for substantial levels of noise (up to $p\approx 0.3$, corresponding to an average of $30\%$ of the edges/non-edges differing between each mode and networks in its cluster). 

The partition distance shows similar gradual performance degradation, with substantial increases in the distance beginning at $p\approx 0.3$ for the contiguous experiments and $p\approx 0.15$ for the unconstrained experiment (Fig.~\ref{fig:recovery}b). The partition distance levels off at different values across the three experiments, with the unconstrained case exhibiting significantly worse performance than the contiguous cases. We expect this result since contiguity simplifies the reconstruction problem by reducing the space of possible clusterings. Because information-based measures account for the entire space of possible clusterings instead of the highly constrained set produced by contiguous partitions, they overestimate the similarity of partitions in this constrained set. This overestimation intensifies with more clusters \cite{kirkley2022spatial}.

The inverse compression ratio (Eq.~\eqref{eq:ratio}) for these experiments gradually approaches $1$ (no compression relative to transmitting each network individually, denoted by the dashed black line) as the noise level $p$ increases (Fig.~\ref{fig:recovery}c). This result is consistent with the intuition that noisier data will be harder to compress, while data with strong internal regularity will be much easier to compress, as the homogeneities can be exploited for shorter encodings. When $p$ is small, we can achieve up to $10$ times compression over the naive baseline by using the inferred underlying modes and clusters to transmit these network populations. 

The results in Fig.~\ref{fig:recovery} indicate that our algorithms can recover the underlying modes and their clusters in synthetic network populations. However, these results also depend on how distinguishable the underlying modes are. For identical modes, $\bm{A}^{(1)}=\bm{A}^{(2)}$, it is impossible to recover the cluster labels of the individual network samples $\bm{D}^{(s)}$. To investigate the dependence of the recovery performance on the modes themselves, we repeat the experiment in Fig.~\ref{fig:recovery}, except this time we systematically vary the mode networks $\mathcal{A}$ for each trial to achieve various levels of distinguishability. In each trial, we set $\bm{A}^{(1)}$ equal to the corresponding mode in Fig.~\ref{fig:diagram} as before, but then generate the edges in $\bm{A}^{(2)}$ from $\bm{A}^{(1)}$ using the flip probability $\gamma$, which we call the ``mode separation''. For mode separations $\gamma\approx 0$, it is challenging to recover the correct cluster labels of the individual sample networks because $\bm{A}^{(2)}$ will closely resemble $\bm{A}^{(1)}$. On the other hand, for mode separations $\gamma\approx 0.5$, the modes will typically be easily distinguished since $\bm{A}^{(2)}$ will have many edges/non-edges that have flipped relative to $\bm{A}^{(1)}$. 

Figure~\ref{fig:recovery2} shows the results of this second experiment. The panels show the partition distance between the true and inferred cluster labels for a range of mode separations $\gamma$. In all experiments, the recovery becomes worse for lower values of the separation $\gamma$, but the algorithm still recovers a significant amount of cluster information even for relatively low $\gamma$. As in the previous set of experiments, the recovery performance is substantially worse for the discontiguous case compared with the contiguous cases, again due to the highly constrained ensemble of possible partitions considered by the partition distance in the contiguous cases. 

In Appendix~\ref{sec:tests}, we show the recovery performance results for the network distance between the true and inferred modes as we vary the mode separation $\gamma$. For the mode recovery, the results are even more robust to the changes in mode separation. This result is consistent with the recovery performance in Fig.~\ref{fig:recovery}, where the recovery performance of the partitions starts to worsen at lower noise levels than the recovery of the modes. Thus, small perturbations in the inferred clusters may not affect the inferred modes much, since misclassified networks likely have little in common with the rest of their cluster.    \\

%%%%%%%%%%%%%%%%%%%%%%%%%%%%%%%%%%%%%%%%%%%%%
%%%%%%%%%%%%%%%%%%%%%%%%%%%%%%%%%%%%%%%%%%%%%

\subsection{Unordered network population representing global trade relationships}
\label{sec:food}

\begin{figure*}
    \centering
    \includegraphics[width=0.8\textwidth]{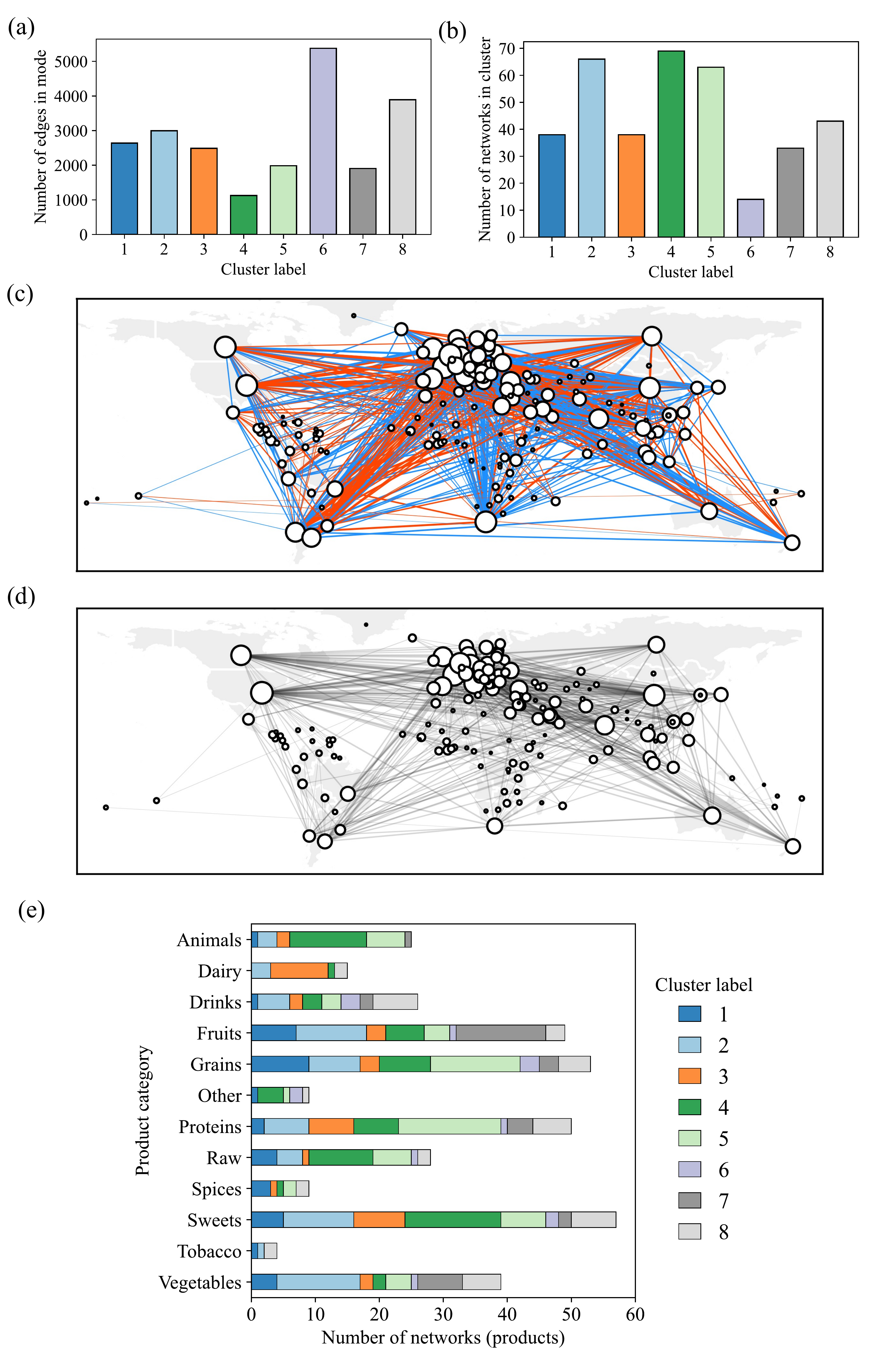}
    \caption{\textbf{Discontiguous networks of imports and exports.} We apply the algorithm of Sec.~\ref{sec:algorithm} to a collection of trade networks~\cite{de2015structural} described in Sec.~\ref{sec:food} to identify similar networks of products. \textbf{(a)} Number of edges in each cluster's mode. \textbf{(b)} Number of networks in each cluster. \textbf{(c)} Edges in mode 7 but not mode 5 are colored in blue, while edges in mode 5 but not in mode 7 are colored in red, highlighting the differences between these two modes. \textbf{(d)} The shared backbone of edges common to both modes 5 and 7. \textbf{(e)} Distribution of product types across the networks in each cluster.}
    \label{fig:fao}
\end{figure*}

For our first example with empirical network data, we study a collection of worldwide import/export networks. The nodes represent countries and the edges encode trading relationships. The Food and Agriculture Organization of the United Nations (FAO) aggregates these data, and we use the trades made in 2010, as in \cite{de2015structural}. Each network in the collection corresponds to a category of products, for example, bread, meat, or cigars. We ignore information about the intensity of trades and merely record the presence or absence of a trading relationship for each category of products. The resulting collection comprises 364 networks (layers) on the same set of 214 nodes with 874.6 edges (average degree of 8.2) on average, with some sparse networks having as little as one edge and the densest containing 6,529 edges. These networks are unordered, so we employ the discontiguous clustering method described in Sec.~\ref{sec:algorithm}. We run the algorithm multiple times with varying initial number of clusters $K_0$ to find the best optima, although as with the synthetic reconstruction examples the choice of $K_0$ has little impact on compression. 
The best compression we find results in 8 modes and achieves a compression ratio of $\eta(\mathcal{D})=0.562$, indicating that it is nearly twice as efficient to communicate the data when we use the modal networks and their clusters.
In contrast, in \cite{de2015structural} a clustering analysis of the same network layers using structural reducibility---a measure of how many layers can be aggregated to reduce pairwise information redundancies among the layers---yielded 182 final aggregated layers, which would poorly compress the data under our scheme and not provide a significant benefit in downstream analyses due to the large final number of clusters.
Key properties of the configuration of modes and clusters inferred by our algorithm are illustrated in Fig.~\ref{fig:fao}. 

In Fig.~\ref{fig:fao}a we show the number of edges in each inferred mode, which indicates that these modes vary substantially in density to reflect the key underlying structures in networks within their corresponding clusters. The sizes of the clusters, shown in Fig.~\ref{fig:fao}b, also vary substantially, with the most populated cluster (cluster 4) containing nearly 7 times as many networks as the least populated cluster (cluster 6). 
Some striking geographical commonalities and differences in the structure of the modes can be seen due to the varying composition of their corresponding clusters of networks. Figures~\ref{fig:fao}c and~\ref{fig:fao}d show the differences and similarities respectively between the structure of the modes for clusters 5 and 7, which are chosen as example modes because of their modest densities and distinct distributions of product types (Fig.~\ref{fig:fao}e). Edges that are in mode 7 but not in mode 5 are highlighted in blue, while edges in mode 5 but not in mode 7 are highlighted in red. Meanwhile, the shared edges common to both networks are shown in Fig.~\ref{fig:fao}d in black. Mode 5, which contains a diversity of product types and a relatively large portion of grain and protein products, has a large number of edges connecting the Americas to Europe that are not present in mode 7. On the other hand, mode 7, which is primarily composed of networks representing the trade of fruits and vegetables, has many edges in the global south that are not present in mode 5. However, both modes share a common backbone of edges that are distributed globally.

We categorized the 364 products (the network layers being clustered) into 12 broader categories of product types, plotting their distributions within each cluster in Fig.~\ref{fig:fao}e. There are a few interesting observations we can make about this figure. Nearly all of the dairy products are traded within networks in a single cluster (cluster 3), indicating a high degree of similarity in the trade patterns for dairy products across countries. A similar observation can be made for live animals, which are primarily traded in cluster 4. On the other hand, many of the other products (grains, proteins, sweets, fruits, vegetables, and drinks) are traded in reasonable proportion in all clusters, which may reflect the diversity of these products as well as their geographical sources, which can give rise to heterogeneous trading structures. The densities of the modes and the sizes of the clusters do not have a clear relationship, with cluster 6 containing the smallest number of networks but the densest mode, and clusters 4 and 5 having sparser modes and much larger clusters. This reflects a higher level of heterogeneity in the structure of the trading relationships captured in cluster 6, which requires a denser mode for optimal compression, while the converse is true for clusters 4 and 5.

We also identify substantial structural differences in the inferred modes. In Appendix~\ref{sec:appendix_modes}, we compute summary statistics (average degree, transitivity, and average betweenness) for the modes output in this experiment and the network layers in their corresponding clusters. The statistics vary much more across clusters than within the clusters, suggesting that the MDL optimal mode configuration exemplifies distinct network structures within the dataset. Because the within-cluster average value of each statistic and the corresponding value for the mode network are similar, our method provides an effective preprocessing step for network-level regression tasks.  

%%%%%%%%%%%%%%%%%%%%%%%%%%%%%%%%%%%%%%%%%%%%%
%%%%%%%%%%%%%%%%%%%%%%%%%%%%%%%%%%%%%%%%%%%%%
\subsection{Ordered network population representing the fossil record}
\label{sec:fossils}

\begin{figure*}
    \centering
    \includegraphics[width=1\textwidth]{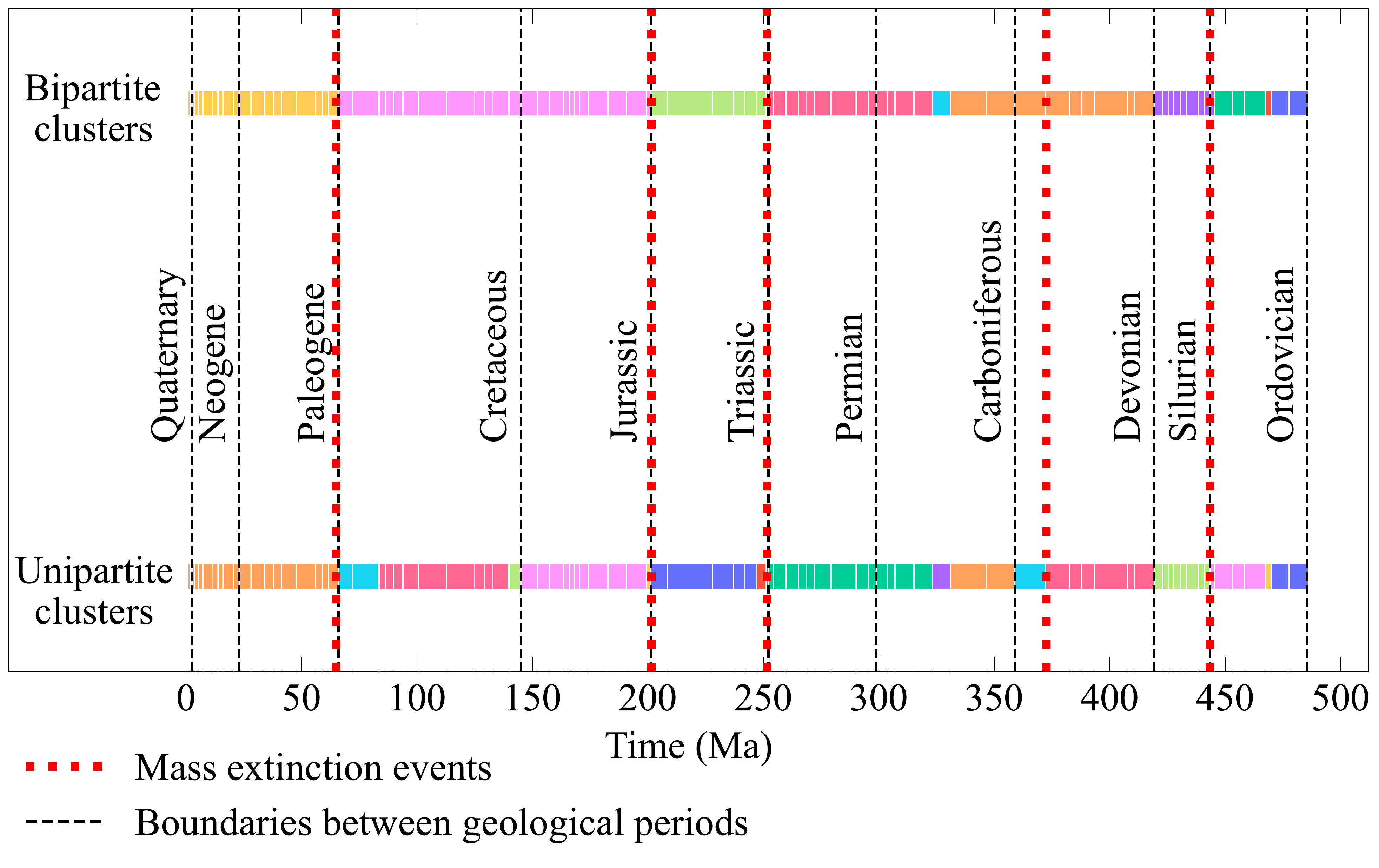}
    \caption{\textbf{Contiguous clusters of network representing the post-Cambrian fossil record.} We apply the dynamic programming algorithm of Sec.~\ref{sec:temporal} to the unipartite genus-genus network population (lower bar) and the bipartite genus-location network population (upper bar) described in Sec.~\ref{sec:fossils} to identify key time intervals with distinct fossil assemblages. The clusters inferred by the algorithm are represented with distinct colors, and the networks, one per each post-Cambrian geological stage, are separated by white lines. Boundaries between geological periods, i.e. larger scale rock units \cite{cohen2013ics}, are indicated by dashed vertical black lines. The five major mass extinction events \cite{raup1982mass} are shown in dotted vertical red lines.}
    \label{fig:fossils}
\end{figure*}

\begin{table*}[htb]
    \centering
    \begin{tabular}{c|c|c|c|c}
        & Eras & Extinction Events &  Periods & MDL Optimal\\
        \hline 
$K(\mathcal{D}_{uni})$ & $3$ & $6$ & $11$ & $17$ \\
\hline 
$K(\mathcal{D}_{bi})$ & $3$ & $6$ & $11$ & $10$ \\
    \hline
    $\eta(\mathcal{D}_{uni})$ & $0.814$ & $0.805$ & $0.810$ & $0.796$\\
    \hline
    $\eta(\mathcal{D}_{bi})$ & $0.842$ & $0.838$ & $0.850$ & $0.833$
    \end{tabular}
    \caption{\textbf{Compression results for different partitions of the fossil record.} $\mathcal{D}_{uni}$ denotes the set of fossil record networks in the unipartite genus-genus representation, and $\mathcal{D}_{bi}$ denotes the same networks in the bipartite genus-location representation.}
    \label{tab:fossils}
\end{table*}

We conclude our analysis with a study of a set of networks representing global marine fauna over the past 500 million years. We aggregate fossil occurrences of the shelled marine animals, including bryozoans, corals, brachiopods, mollusks, arthropods, and echinoderms, into a regular grid covering the Earth’s surface \cite{rojas_multiscale_2021}. From these data, we construct unweighted bipartite networks representing $90$ ordered time intervals in Earth's history (geological stages): An edge between a genus and a grid cell indicates that the genus was observed in the grid cell during the network's corresponding geological stage. We also construct the unipartite projections of these networks: An edge from one genus to another indicates that these two genera were present in the same grid cell during the stage corresponding to the network. In total, there were 18,297 genus nodes, 664 grid cell nodes, 67,371.5 edges on average for the 90 unipartite graphs (average degree of 7.4), and 1462.2 edges on average for the bipartite graphs (average degree of 0.08, corresponding to an average of roughly 10 percent of genera being present at each layer). 

In Fig.~\ref{fig:fossils}, we show the results of applying our clustering method for contiguous network populations (Sec.~\ref{sec:temporal}) to both the unipartite and bipartite populations representing the post-Cambrian fossil record. We find clusters that capture the known large-scale organization of marine diversity. Major groups of marine animals archived in the fossil record are organized into global-scale assemblages that sequentially dominated oceans and shifted across major biotic transitions. Overall, the bipartite and unipartite fossil record network representations both result in transitions concurrent with the major known geological perturbations in Earth’s history, including the so-called mass extinction events. However, differences in the clusters retrieved from the unipartite and bipartite representations of the underlying paleontological data highlight the impact of this choice on the observed macroevolutionary pattern \cite{eriksson_how_2021}. 

We also use our methodology to assess the extent to which the standard division of the post-Cambrian rock record in the geological time scale and the well-known mass extinction events compress the assembled networks. Specifically, we evaluate the inverse compression ratio in Eq.~\eqref{eq:ratio} on three different partitions of the fossil record networks that are defined by clustering the assembled networks into geological eras (Paleozoic, Mesozoic, and Cenozoic), geological periods (Ordovician to Quaternary), and six time intervals between the five mass extinctions in Fig.~\ref{fig:fossils}, with planted modes constructed by placing the networks into each cluster using the algorithm described in Sec.~\ref{sec:algorithm} and Appendix~\ref{sec:appendixmergesplit}.

Table~\ref{tab:fossils} shows the results of these experiments. All three partitions compress the fossil record networks almost as much as the optimal partition, which represents a natural division based on major regularities. Accordingly, the planted partition based on mass extinctions is almost as good as this optimal partition because mass extinctions are concurrent with the major geological events shaping the history of marine life. In contrast, partitions based either on standard geological eras or periods are less optimal, likely because they represent, to some extent, arbitrary divisions that are maintained for historical reasons. Our results here provide a complementary perspective to the work in \cite{rojas_multiscale_2021}, where a multilayer network clustering algorithm was employed that clusters nodes within and across layers to reveal three major biotic transitions from the fossil data. In Appendix~\ref{sec:comparison} we review this and other existing multiplex and network population clustering techniques, discussing the similarities and differences with our proposed methods.

%%%%%%%%%%%%%%%%%%%%%%%%%%%%%%%%%%%%%%%%%%%%%
%%%%%%%%%%%%%%%%%%%%%%%%%%%%%%%%%%%%%%%%%%%%%

\section{Conclusion}
  
We have used the minimum description length principle to develop efficient parameter-free methods for summarizing populations of networks using a small set of representative modal networks that succinctly describe the variation across the population. For clustering network populations with no ordering, we have developed a fast merge-split Monte Carlo procedure that performs a series of moves to refine a partition of the networks. For clustering ordered networks into contiguous clusters, we employ a time and memory-efficient dynamic programming approach. These algorithms can accurately reconstruct modes and associated clusters in synthetic datasets and identify significant heterogeneities in real network datasets derived from trading relationships and fossil records. Our methods are principled, non-parametric, and efficient in summarizing complex sets of independent network measurements, providing an essential tool for exploratory and visual analyses of network data and preprocessing large sets of network measurements for downstream applications.  

This information-theoretic framework for representing network populations with modal networks can be extended in several ways. For example, a multi-step encoding that allows for hierarchical partitions of network populations would capture multiple levels of heterogeneity in the data. More complex encodings that exploit structural regularities within the networks would allow for simultaneous inference of mesoscale structures---such as communities, core-periphery divisions, or specific informative subgraphs \cite{vreeken2022differential}---along with the modes and clusters. The encodings can also be adapted to capture weighted networks with multiedges by altering the combinatorial expressions for the number of allowable edge configurations.

%%%%%%%%%%%%%%%%%%%%%%%%%%%%%%%%%%%%%%%%%%%%%
%%%%%%%%%%%%%%%%%%%%%%%%%%%%%%%%%%%%%%%%%%%%%

\section*{Acknowledgments}
\vspace{-\baselineskip}
A.K. was supported in part by the HKU-100 Start Up Grant. M.R. was supported by the Swedish Research Council, Grant No.\ 2016-00796.

%%%%%%%%%%%%%%%%%%%%%%%%%%%%%%%%%%%%%%%%%%%%%
%%%%%%%%%%%%%%%%%%%%%%%%%%%%%%%%%%%%%%%%%%%%%

\section{Code Availability}

The algorithm presented in this paper is available at \url{https://github.com/aleckirkley/MDL-network-population-clustering}.

%%%%%%%%%%%%%%%%%%%%%%%%%%%%%%%%%%%%%%%%%%%%%
%%%%%%%%%%%%%%%%%%%%%%%%%%%%%%%%%%%%%%%%%%%%%

\section{Data Availability}

The data sets used in this paper are available at \url{https://github.com/aleckirkley/MDL-network-population-clustering}.

%%%%%%%%%%%%%%%%%%%%%%%%%%%%%%%%%%%%%%%%%%%%%
%%%%%%%%%%%%%%%%%%%%%%%%%%%%%%%%%%%%%%%%%%%%%

\section{Author Contributions}

A.K. designed the study and methodology, A.K., A.R., M.R., and J.G.Y. designed the experiments, A.K. and J.G.Y. performed the experiments, A.R. and M.R provided new datasets, and A.K. wrote the manuscript. All authors reviewed, edited and approved the manuscript.

%%%%%%%%%%%%%%%%%%%%%%%%%%%%%%%%%%%%%%%%%%%%%
%%%%%%%%%%%%%%%%%%%%%%%%%%%%%%%%%%%%%%%%%%%%%

\section{Competing Interests}

The authors declare no competing interests.

%%%%%%%%%%%%%%%%%%%%%%%%%%%%%%%%%%%%%%%%%%%%%
%%%%%%%%%%%%%%%%%%%%%%%%%%%%%%%%%%%%%%%%%%%%%

%%%%%%%%%%%%%%%%%%%%%%%%%%%%%%%%%%%%%%%%%%%%%
%%%%%%%%%%%%%%%%%%%%%%%%%%%%%%%%%%%%%%%%%%%%%

\appendix
\onecolumngrid

\section{Merge-split Monte Carlo algorithm}
\label{sec:appendixmergesplit}

In this appendix we describe the greedy merge-split Monte Carlo algorithm used to identify the clusters $\mathcal{C}$ and modes $\mathcal{A}$ that minimize the description length in Eq.~\eqref{eq:DLtotal}. The input to the algorithm is $S$ edge sets corresponding to the network population $\mathcal{D}$, an initial number of clusters $K_0$, and a maximum number of consecutive failed moves $n_{\mathrm{fails}}$ before termination. ($n_{\mathrm{fails}}=100$ for all experiments performed in the paper.) Note that the number of true-positive edges, which we denote $t_k=S_kM_k-n_k$, will give equivalent results if swapped for $n_k$ in Eq.~\ref{eq:Lk}. We will use these true-positive counts $t_k$ in the algorithm below rather than the false-negatives $n_k$, keeping in mind that they are interchangeable in the description length expressions. 

The algorithm is as follows:
\begin{enumerate}

    \item Initialize $K_0$ clusters $\mathcal{C}$ by placing each network in $\mathcal{D}$ in a cluster chosen uniformly at random.
    
    \item For all clusters $C_k$, create a dictionary $E_k=\{(i,j):X^{(k)}_{ij}\}$ mapping all edges $(i,j)$ in networks within $C_k$ to the number of times $(i,j)$ occurs in $C_k$. We do not need to keep track of the edges that never occur in a cluster. This is one computational bottleneck of the algorithm, requiring $O(SN)$ operations for sparse networks with $O(N)$ edges each.
    
    \item For all clusters $C_k$, compute the mode $\bm{A}^{(k)}$ using the following greedy edge removal algorithm:
    
    \begin{enumerate}
    
        \item Create list of tuples $\tilde E_k=\{((i,j),X^{(k)}_{ij}):(i,j)\in E_k\}$, with items sorted in increasing order of frequency $X^{(k)}_{ij}$. This is the computational bottleneck of this mode update step, with roughly $O(N\log N)$ time complexity assuming the networks in cluster $C_k$ have $O(N)$ unique edges. 
    
        \item Initialize:
        \begin{itemize}
            \item $r\leftarrow 0$
            \item $t_k\leftarrow \sum_{(i,j)\in C_k}X^{(k)}_{ij}$
            \item $p_k\leftarrow 0$
            \item $M_k\leftarrow \abs{\tilde E_k}$
            \item $\bm{A}^{(k)}\leftarrow \{(i,j):(i,j)\in E_k\}$
            \item $\Delta \mathcal{L} \leftarrow 0$
             \item $\Delta \mathcal{L}_{best} \leftarrow 0$
             \item $r_{best} \leftarrow 0$
        \end{itemize}
        
        \item While $r<\abs{\tilde E_k}$:
        \begin{enumerate}
            \item $(i,j),X_{ij}^{(k)}\leftarrow \tilde E_k[r]$
            \item Compute the change in description length 
            \begin{align}
        ~~~~~~~~~~~~~~~~~~~\Delta     \mathcal{L}_{sub}&=\mathcal{L}(\bm{A}^{(k)}\setminus \{(i,j)\},C_k) - \mathcal{L}(\bm{A}^{(k)},C_k).   
            \end{align}
            The first term is identical to the second, but with $t_k\to t_k-X^{(k)}_{ij}$, $p_k\to p_k+X^{(k)}_{ij}$, and $M_k\to M_k-1$. 
            
            \item $\Delta \mathcal{L} \leftarrow \Delta \mathcal{L} + \Delta \mathcal{L}_{sub}$
            
            \item $\bm{A}^{(k)} \leftarrow \bm{A}^{(k)}\setminus \{(i,j)\}$ 
                \item Increment:
                \begin{itemize}
                    \item $r\leftarrow r+1$
                    \item $t_k\leftarrow t_k-X^{(k)}_{ij}$
                    \item $p_k\leftarrow p_k+X^{(k)}_{ij}$
                    \item $M_k\leftarrow M_k-1$
                \end{itemize}
                
            \item If $\Delta \mathcal{L}<\Delta \mathcal{L}_{best}$:
            \begin{itemize}
                \item $\Delta \mathcal{L}_{best} \leftarrow \Delta \mathcal{L}$
                \item $r_{best} \leftarrow r$
            \end{itemize}
    
        \end{enumerate}
        
        \item return $\bm{A}^{(k)}$ with the first $r_{best}$ edges in $\tilde E_k$ removed
        
    \end{enumerate}
    
    \item Initialize counter $n_{f}\leftarrow 0$ and compute the total description length $\mathcal{L}(\mathcal{D})$ for the current cluster and mode configurations using Eq.~\eqref{eq:DLtotal}.
    
    \item While $n_f<n_{\mathrm{fails}}$, attempt to do one of the following four moves, chosen at random:
    
        \begin{enumerate}
        
            \item Move 1: Reassign a randomly chosen network $\bm{D}^{(s)}$ from its cluster $C_k$ to the cluster $C_{k'}$ that most reduces the description length. 
            \begin{enumerate}
            
              \item For all $k'\neq k$, compute
                \begin{align}
                ~~~~~~~~~~~~~~~~~~\Delta\mathcal{L}_1(k') &= \mathcal{L}(\bm{A}^{(k')},C_{k'}\cup\{\bm{D}^{(s)}\}) 
                + \mathcal{L}(\bm{A}^{(k)},C_k\setminus \{\bm{D}^{(s)}\})
                -\mathcal{L}(\bm{A}^{(k')},C_{k'}) 
                -\mathcal{L}(\bm{A}^{(k)},C_k).\nonumber
                \end{align}
                The modes are not updated to account for the inclusion/absence of $\bm{D}^{(s)}$, and so we only require updates to $E_k$ and $E_{k'}$ to account for this reassignment, in order to compute new values of $t_k,p_k,t_{k'},p_{k'}$. These edge dictionary updates can be done in $O(N)$ operations for sparse graphs.
                
                \item If $\Delta\mathcal{L}_1(k'')<0$ for some $k''\neq k$:
                \begin{itemize}
                    \item Move $s$ to the cluster $C_{k'}$ that minimizes $\Delta\mathcal{L}_1(k')$, and remove $s$ from $C_k$
                    \item $n_f\leftarrow 0$
                    \item Update $E_k,~E_{k'}$ to account for the new absence/presence of the edges in $\bm{D}^{(s)}$
                    \item Update modes $\bm{A}^{(k)}$ and $\bm{A}^{(k')}$ using the new dictionaries $E_k,~E_{k'}$
                    \item Update the description length with the corresponding change after the move.
                \end{itemize}
                
                \item Else: $n_f \leftarrow n_f+1$ 
                
                \item Time complexity: $O(KN)$ for sparse networks, as computing the change in the description length for the move to $k'$ requires the $O(N)$ operation of comparing the edge list of $\bm{D}^{(s)}$ with the edge list of $\bm{A}^{(k')}$.  
            \end{enumerate}

            \item Move 2: Merge a randomly chosen pair of clusters $C_{k'}$ and $C_{k''}$ to form a new cluster $C_{k}=C_{k'}\cup C_{k''}$.
            \begin{enumerate}
                \item Compute the dictionary $E_k$ by merging the counts in the dictionaries $E_{k'}$ and $E_{k''}$.
                \item Compute $\bm{A}^{(k)}$ using the method outlined in Step 3 of the algorithm. 
                \item Compute the change in description length
                \begin{align}
                 ~~~~~~~~~~~~~~~~\Delta\mathcal{L}_2&=\mathcal{L}(\bm{A}^{(k)},C_{k})-\mathcal{L}(\bm{A}^{(k')},C_{k'}) 
                -\mathcal{L}(\bm{A}^{(k'')},C_{k''}).    
                \end{align}
                \item If $\Delta \mathcal{L}_2<0$:
                \begin{itemize}
                    \item Add $C_{k}$ to $\mathcal{C}$ and delete $C_{k''}$ and $C_{k''}$
                    \item Keep $E_k$ and delete $E_{k'}$ and $E_{k''}$
                    \item Keep $\bm{A}^{(k)}$ and delete $\bm{A}^{(k')}$ and $\bm{A}^{(k'')}$
                    \item $\mathcal{L}(\mathcal{D}) \leftarrow \mathcal{L}(\mathcal{D})+\Delta \mathcal{L}_2$
                \end{itemize}
                
                \item Else: $n_f\leftarrow n_f+1$ 
                \item Time complexity: roughly $O(N\log N)$, with the bottleneck being the computation of the mode $\bm{A}^{(k)}$. 
            \end{enumerate}

            \item Move 3: Split a randomly chosen cluster $C_{k}$ to form new clusters $C_{k'}$ and $C_{k''}$.
            \begin{enumerate}
                \item Initialize two clusters and their edge dictionaries using the method in Steps 1 and 2 of the algorithm.
                \item Identify the modes of $C_{k'}$ and $C_{k''}$ using the method in Step 3 of the algorithm.
                \item Run a $2$-means style algorithm to compute $C_{k'}$ and $C_{k''}$ and their corresponding modes, which proceeds as follows: While networks are still reassigned at the current move:
                \begin{enumerate}
                    \item For all networks $\bm{D}^{(s)}$ in both sub-clusters, compute $\Delta\mathcal{L}_1(k')$ and $\Delta\mathcal{L}_1(k'')$, and add $\bm{D}^{(s)}$ to the cluster with the lower of the two values.
                    \item Update $E_{k'}$ and $E_{k''}$ using the method in Step 2 of the overall algorithm.
                    \item Update the modes of $C_{k'}$ and $C_{k''}$ using the method in Step 3 of the overall algorithm. 
                \end{enumerate}
                \item Compute the change in description length
                \begin{align}
                ~~~~~~~~~~~~~~~\Delta\mathcal{L}_3&=\mathcal{L}(\bm{A}^{(k')},C_{k'})+\mathcal{L}(\bm{A}^{(k'')},C_{k''}) 
                -\mathcal{L}(\bm{A}^{(k)},C_{k}).    
                \end{align}
                
                \item If $\Delta \mathcal{L}_3<0$:
                \begin{itemize}
                    \item Add $C_{k'}$ and $C_{k''}$ to $\mathcal{C}$, delete $C_{k}$
                    \item Keep $E_{k'}$ and $E_{k''}$, delete $E_k$
                    \item Keep $\bm{A}^{(k')}$ and $\bm{A}^{(k'')}$, delete $\bm{A}^{(k)}$
                    \item $\mathcal{L}(\mathcal{D}) \leftarrow \mathcal{L}(\mathcal{D})+\Delta \mathcal{L}_3$  
                \end{itemize}

                \item Else: $n_f\leftarrow n_f + 1$
                
                \item Time complexity: $O(NS/K)$ for roughly equally sized clusters, with the computational bottleneck being the reassignment step in the local $2$-means algorithm.
            \end{enumerate}
            
            \item Move 4: Pick two clusters at random to merge, then immediately split following Move 3. 
            \begin{itemize}
                \item Time complexity: $O(NS/K)$, similar to Move 3.
            \end{itemize}

        \end{enumerate}

\end{enumerate}

Now that we have laid out the merge-split algorithm, we can show that the greedy mode update in Step 3 of the merge-split algorithm is locally optimal under a few assumptions regarding the sparsity of the networks and the set of edges in the cluster as a whole. The relevant terms of the local description length that we need to consider are those that depend on the choice of mode, or
\begin{align}
\label{eq:phi}
\phi_k(\bm{A}^{(k)},C_k)&\approx \log {{N\choose 2}\choose M_k}+\log {S_kM_k \choose t_k}+\log {S_k{N\choose 2}\choose p_k},
\end{align}
were we've approximated $ M_k^\ast ={N\choose 2}-M_k\approx {N\choose 2}$, which will be a good approximation when the modes are sparse. We also note the use of true-positives $t_k$ rather than false-negatives $n_k$, as discussed at the beginning of the section.

We can first notice that in the sparse network regime, it will never be optimal to include an edge $(i,j)$ in the mode $\bm{A}^{(k)}$ that does not exist in any of the networks $\bm{D}^{(s)}\in C_k$. If $M_k <\frac{1}{2}{N\choose 2}-1$, then including an edge $(i,j)$ in the mode $\bm{A}^{(k)}$ that does not exist in any of the networks $\bm{D}^{(s)}\in C_k$ will only serve to augment $M_k$ by $1$, increasing the values of the first two terms of $\phi_k$ in Eq.~\eqref{eq:phi}. Thus, we will end up increasing the description length if we add such an edge, and so no such edges will exist in the optimal mode $\bm{A}^{(k)}$. We can therefore begin our algorithm with the maximally dense mode $\bm{A}_{\comp}^{(k)}$ in Eq.~\eqref{eq:Acomp} and only consider removing edges, since the optimal mode will be comprised of some subset of these edges. 

We can now argue that, for any current iteration of the mode, the best edge $(i,j)$ to remove from the mode next is the one that occurs the least often in the cluster $C_k$ (i.e. the number of this edge's occurrences, $X^{(k)}_{ij}$, is the smallest). $M_k$ will always decrease by $1$ regardless of which edge is chosen, but $t_k$ and $p_k$ will decrease and increase by $X^{(k)}_{ij}$ respectively. If we assume that $t_k/S_k>M_k/2$---on average, each network in $C_k$ has at least half of the edges in the mode $\bm{A}^{(k)}$, which will be true when the networks in $C_k$ have relatively high overlap in their edge sets---then the second term in Eq.~\eqref{eq:phi} will increase its value by the smallest amount under this edge removal when $t_k$ decreases by the least amount. Likewise, the third term in Eq.~\eqref{eq:phi} will also increase its value with this edge removal (since $p_k$ increases by $X^{(k)}_{ij}$, and we assume that $p_k\ll S_k{N\choose 2}$), and so it will increase by the smallest amount when $X^{(k)}_{ij}$ is smallest. The best edge $(i,j)$ to remove from the mode in any given step of the algorithm is, therefore, the one with the smallest value of $X^{(k)}_{ij}$, as it results in the most negative change in $\phi_k$. Therefore the greedy edge removal algorithm will reach a locally optimal edge configuration for the mode under these assumptions. 

In practice, the total run time of this merge-split algorithm depends on the user's choice of $n_{\mathrm{fails}}$, the number of consecutive failed moves before algorithm termination. When computation time is not a concern, it is safer to let $n_{\mathrm{fails}}$ be as high as possible to ensure no additional merge-split moves will increase the description length. For context, the convergence of the algorithm in the global trade experiment presented here with $n_{\mathrm{fails}}=100$ required on average $150$ merge-split steps and a total run time of $15$ minutes for a pure Python implementation with no substantial optimizations on an Intel Core i7 CPU. On the other hand, with $n_{\mathrm{fails}}=10$ this same experiment takes on average $36.4$ merge-split steps and a total run time of $3.7$ minutes on the same processor. In the synthetic network experiments, each simulation took on average $0.39$ seconds for $138$ MCMC moves with $n_{\mathrm{fails}}=100$ ($S$ and $N$ are much smaller, resulting in shorter run times). The contiguous clustering algorithm does not present the same variability in total run time since it is deterministic given the input networks.

%%%%%%%%%%%%%%%%%%%%%%%%%%%%%%%%%%%%%%%%%%%%%
%%%%%%%%%%%%%%%%%%%%%%%%%%%%%%%%%%%%%%%%%%%%%

\section{Additional synthetic reconstruction tests}
\label{sec:tests}

In this appendix we show the results of additional synthetic reconstruction results for the discontiguous and contiguous network population clustering algorithms of Sec.s~\ref{sec:algorithm} and Sec.~\ref{sec:temporal}. These results verify the robustness of these algorithms as well as their theoretical run time scaling.  

\begin{figure}
    \centering
    \includegraphics[width=0.5\textwidth]{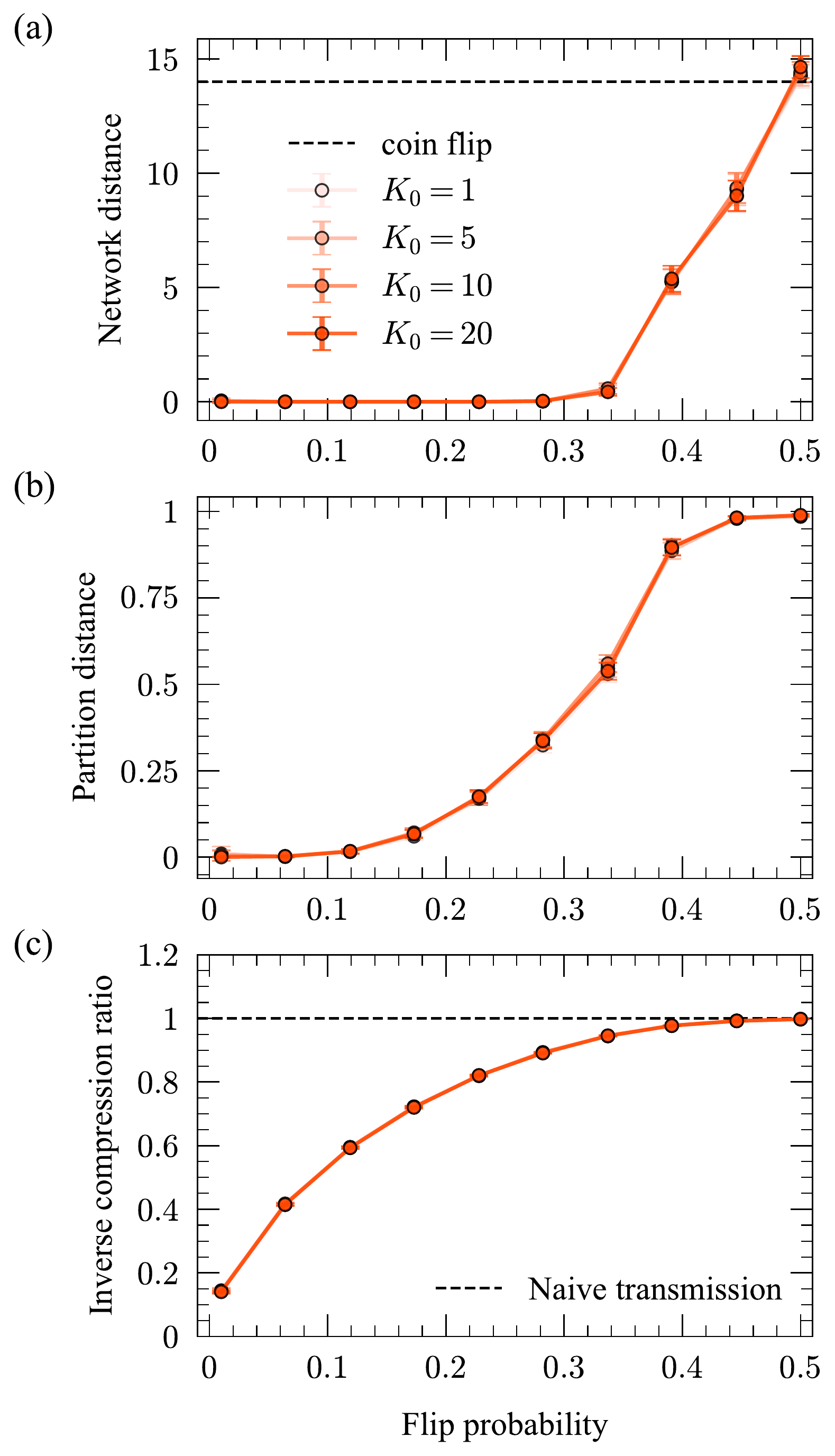}
    \caption{\textbf{Cluster recovery for different initial numbers of clusters $K_0$}. Recovery performance is plotted for the three experiments described in Sec.~\ref{sec:synthetic}, for different initial numbers of clusters $K_0$. \textbf{(a)} Network distance, for various flip probabilities $p$. \textbf{(b)} Partition distance. \textbf{(c)} Inverse compression ratio (Eq.~\eqref{eq:ratio}). Each data point is an average over $200$ realizations of the population for the corresponding value of the flip probability, and error bars correspond to three standard errors in the mean. }
    \label{fig:K0recovery}
\end{figure}

\begin{figure}
    \centering
    \includegraphics[width=0.5\textwidth]{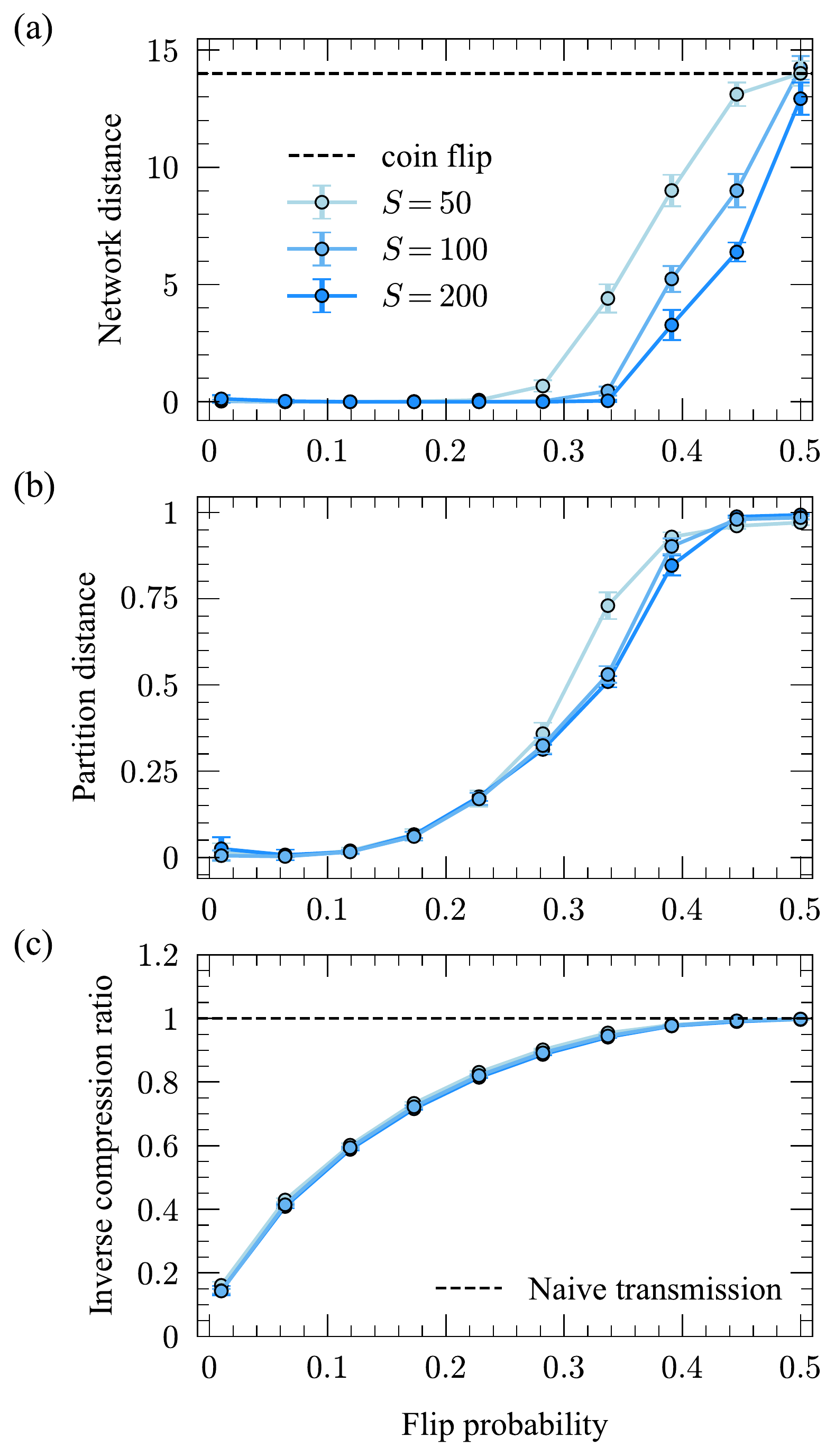}
    \caption{\textbf{Cluster recovery for different population sizes $S$}. Recovery performance is plotted for the three experiments described in Sec.~\ref{sec:synthetic}, for different population sizes $S$. \textbf{(a)} Network distance, for various flip probabilities $p$. \textbf{(b)} Partition distance. \textbf{(c)} Inverse compression ratio (Eq.~\eqref{eq:ratio}). Each data point is an average over $200$ realizations of the population for the corresponding value of the flip probability, and error bars correspond to three standard errors in the mean. }
    \label{fig:Srecovery}
\end{figure}

In Fig.~\ref{fig:K0recovery}, we plot the results shown in Fig.~\ref{fig:recovery} for different choices of the initial number of clusters $K_0$ in the discontiguous case (the contiguous algorithm always begins with all networks in their own cluster), finding that the results are practically indistinguishable for all choices $K_0=1,5,10,20$. In Fig.~\ref{fig:Srecovery}, we repeat the same experiment but this time for different population sizes $S$, finding that the results do depend on $S$, but in an intuitive way: more networks give us more statistical evidence for the clusters and their modes, resulting in better reconstruction.    

\begin{figure}
    \centering
    \includegraphics[width=0.45\textwidth]{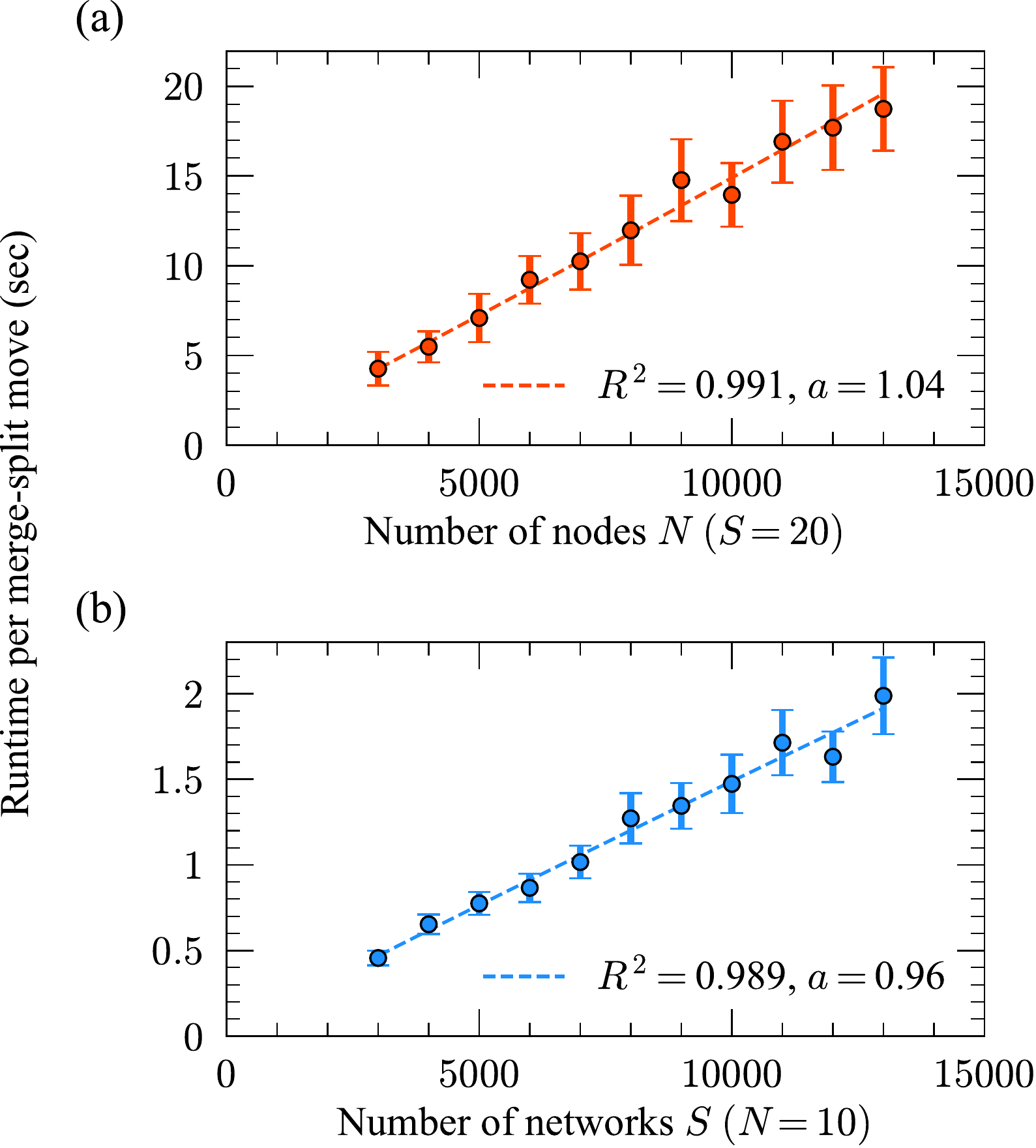}
    \caption{\textbf{Runtime scaling for discontiguous  clustering MCMC moves, for different network sizes $N$ and population sizes $S$}. In each realization a synthetic population was generated using the procedure outlined in Appendix~\ref{sec:tests}, and the average runtime for the merge-split move (Move 4 in Appendix~\ref{sec:appendixmergesplit}, which is the computational bottleneck for $K>1$) was recorded when applying the algorithm of Sec.~\ref{sec:algorithm} to the synthetic data. Regressions $\log y =a\log x + b$ were run for each experiment (dashed lines), and the resulting slopes and $R^2$ values are shown in the corresponding figure panels. Each data point is an average over $100$ synthetic realizations, error bars corresponding to three standard errors.}
    \label{fig:scaling_discontiguous}
\end{figure}

In Fig.~\ref{fig:scaling_discontiguous}, we numerically verify the runtime scaling for the MCMC algorithm discussed in Sec.~\ref{sec:algorithm} and Appendix~\ref{sec:appendixmergesplit}. For each realization of the experiment, we first constructed two Erdos-Renyi random graphs with $N$ nodes and average degree $5$---this value does not matter for the scaling so long as it is constant across $N,S$---which served as modes for the synthetic population. We then generate a population of $S$ networks from these modes using the model of Appendix~\ref{sec:MAP}, each network having a 50/50 chance of being generated from either mode, with true-positive rates $\alpha_1=\alpha_2=0.9$ and false positive rates $\beta_1=\beta_2=0$ to introduce noise into the population but facilitate relatively simple recovery. The MCMC algorithm of Sec.~\ref{sec:algorithm} was then applied to each population, and the average run time of Move 4 in the algorithm was recorded---this move is the computational bottleneck for $K>1$ since it involves combining a merge and a split move (Moves 2 and 3). Fig.~\ref{fig:scaling_discontiguous} shows the average run time over 100 realizations of synthetic populations for a range of $N$ values (at fixed $S$) and range of $S$ values (at fixed $N$), with error bars corresponding to three standard errors in the mean. The regression $\log y=a\log x+b$ was run for the experiments in each panel, and the resulting slopes $a\approx 1$ and coefficients of determination $R^2$ provide strong numerical evidence that the theoretical run time scaling $O(NS)$ is obeyed for this algorithm. The dependence on $K$ is not investigated since the algorithm is constantly updating this parameter. However, at $K=1$ (for Move 3) and $K=2$ (for Moves 2 and 4) the slowest MCMC moves will occur since the largest clusters need to be dealt with, and these will contribute the largest run times to the scaling so in practice the $O(NS)$ dependence is the primary one of interest. 

\begin{figure}
    \centering
    \includegraphics[width=0.45\textwidth]{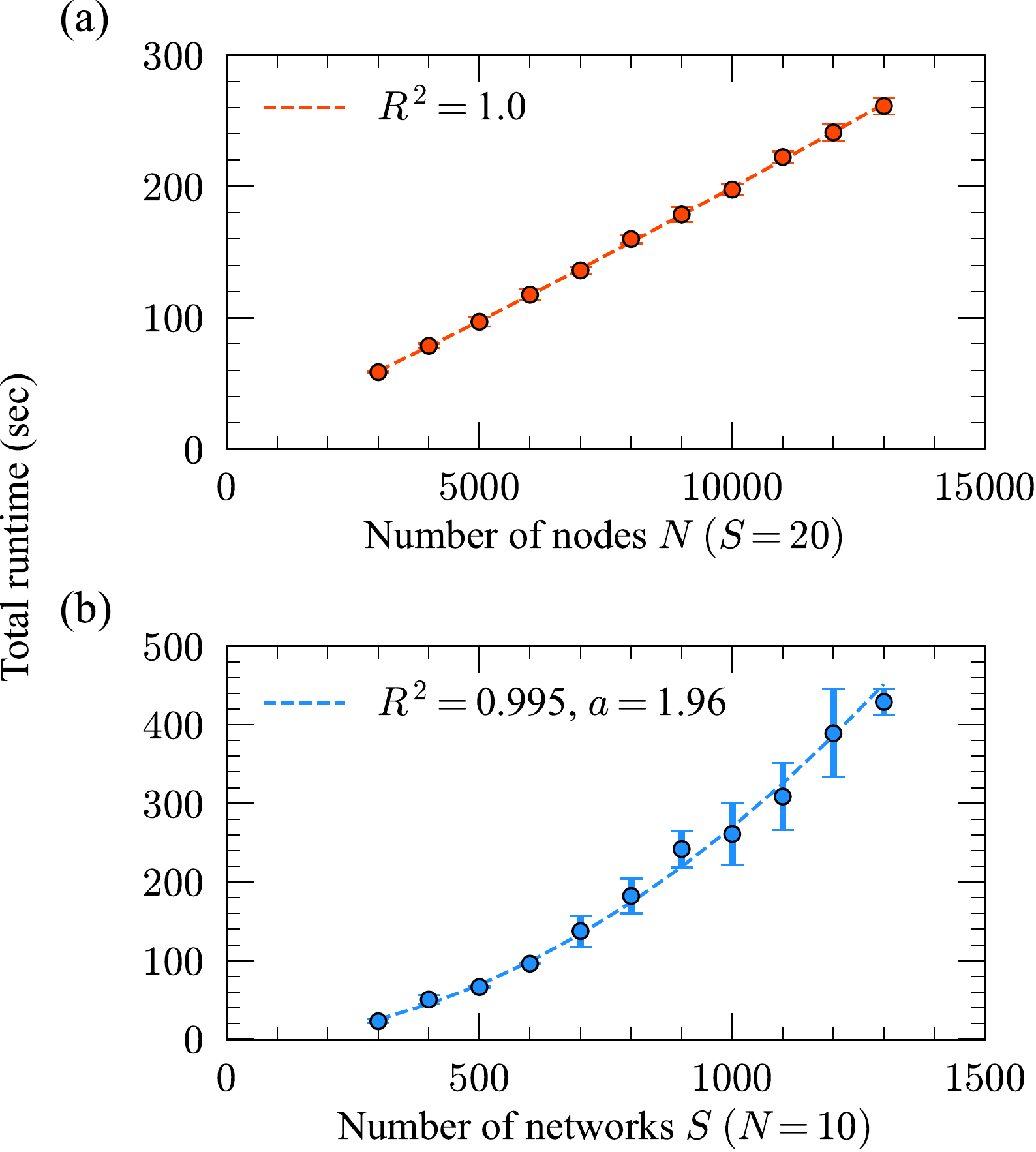}
    \caption{\textbf{Runtime scaling for the  contiguous clustering dynamic program, for different network sizes $N$ and population sizes $S$}. In each realization a synthetic population was generated using the procedure outlined in Appendix~\ref{sec:tests}, and the total runtime was recorded when applying the dynamic programming algorithm of Sec.~\ref{sec:temporal} to the synthetic data. Regressions $y =a(x\log x)$ and $\log y = a \log x + b$ were run for the top and bottom panels respectively (dashed lines), and the resulting slopes and $R^2$ values are shown in each panel. Each data point is an average over $3$ synthetic realizations (each realization showed remarkable runtime consistency), errorbars corresponding to three standard errors.}
    \label{fig:scaling_contiguous}
\end{figure}

A similar experiment was run for synthetic contiguous populations with two planted clusters to verify the theoretical run time scaling for the dynamic programming algorithm. The modes for each realization were generated in the same way as for the previous experiment, except this time the first $S/2$ networks were generated from the first mode and the final $S/2$ networks were generated from the second mode. The dynamic programming of Sec.~\ref{sec:temporal} was then applied to each population, and the average run time of the full algorithm was computed. Fig.~\ref{fig:scaling_contiguous} shows the average run time over 3 realizations of synthetic populations (each run gave a very consistent run time) for a range of $N$ values (at fixed $S$) and range of $S$ values (at fixed $N$), with error bars corresponding to three standard errors in the mean. The regression $ y=a(x\log x)$ was run for the experiment in panel (a) and the regression $\log y=a\log x+b$ was run for the experiment in panel (b). The resulting coefficients of determination $R^2$ and slope $a\approx 2$ for the second regression provide strong numerical evidence that the theoretical run time scaling $O(S^2N\log N)$ is obeyed for this algorithm.

\begin{figure}
    \centering
    \includegraphics[width=0.5\columnwidth]{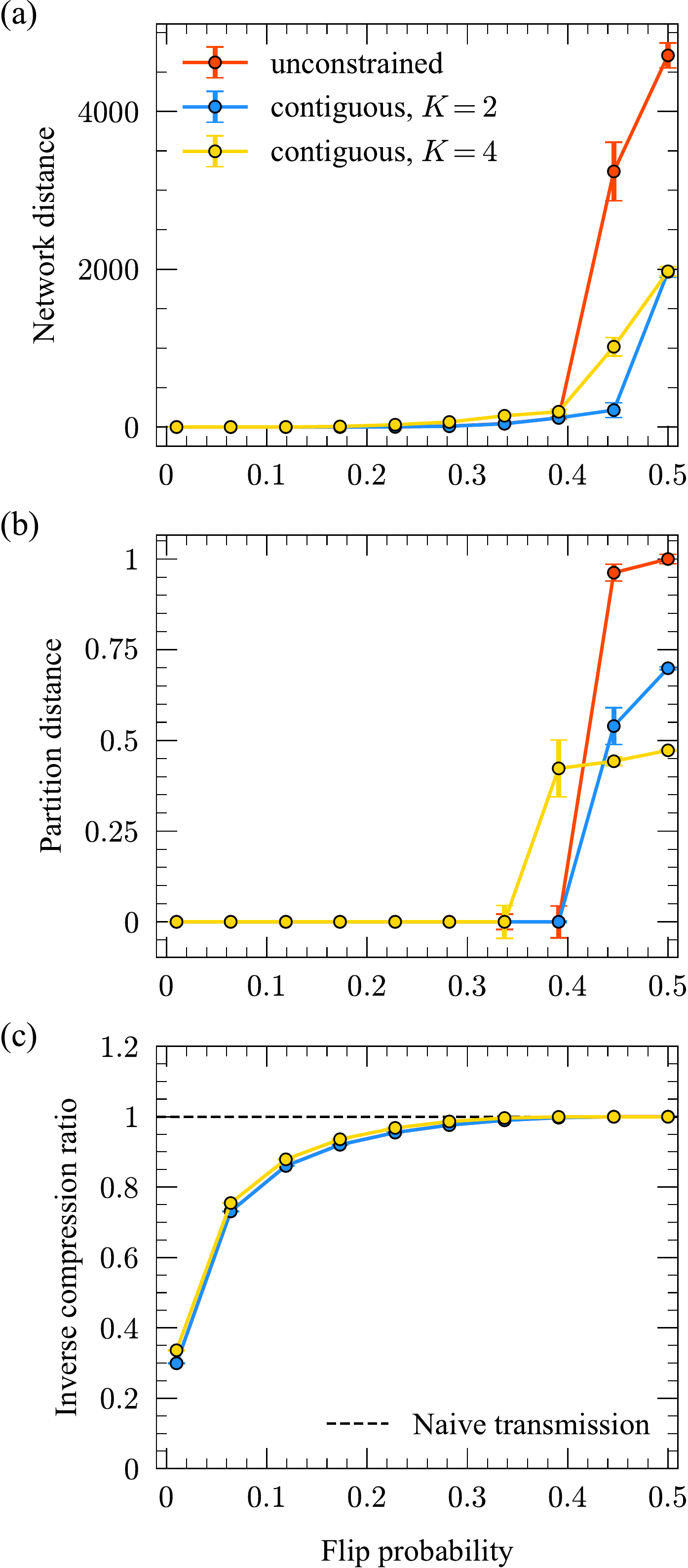}
    \caption{\textbf{Reconstruction experiments with larger modes}. Same reconstruction experiments of Fig.~\ref{fig:recovery}, except with the two modes generated from an Erdos-Renyi random graph with $N=100$ nodes and average degree $c=5$.}
    \label{fig:recoveryN100}
\end{figure}

In Fig.~\ref{fig:recoveryN100}, we rerun the reconstruction experiments of Fig.~\ref{fig:recovery} with the same parameters except with the two modes generated from an Erdos-Renyi random graph with $N=100$ nodes and average degree $c=5$ as in the previous experiments. In general, the recovery performance remains qualitatively similar to that of Fig.~\ref{fig:recovery}, except with variations in the high noise regime ($p\sim 0.5$) due to size-dependent effects from the synthetic population generative process. In this regime, there is a very low probability of any substantial overlap among the networks in the population due to the quadratic scaling of the number of possible positions where an edge can be placed. The discontiguous clustering method (red curves) places all networks in a single cluster which is the most parsimonious description of an unstructured population. This outcome results in one nearly complete inferred mode of $E\sim {N\choose 2}$ edges. By the symmetry of the binomial coefficient, it is equally economical in an information-theoretic sense to specify positions of non-edges than edges. Considering the density of the true modes, we then have network hamming distances of roughly ${N\choose 2}-cN/2\sim 4700$, consistent with the red curve in the top panel of the figure.

In the same $p\sim 0.5$ regime for the $N=100$ case, the contiguous clustering methods tend to do the opposite and place all networks in their own cluster since the corresponding description length does not incur as strong a penalty for extra clusters. Here we then see inferred modes that consist of the individual networks in the population---in other words, random coin flips at each edge position with half of the ${N\choose 2}$ possible edges occupied. Considering the expected overlap with the true modes and fluctuations, we get mode hamming distances of roughly $2000$, as seen in the top panel. Meanwhile, as opposed to the $K=1$ discontiguous case where the NMI equals $0$ (i.e. subtracting the NMI from $1$ gives a partition distance of $1$), in the contiguous setting we see NMI values of $2\log K/(\log S+\log K)$, resulting in partition distances of $0.70,0.48$ for $K=2,4$ respectively.

\begin{figure}
    \centering
    \includegraphics[width=0.5\columnwidth]{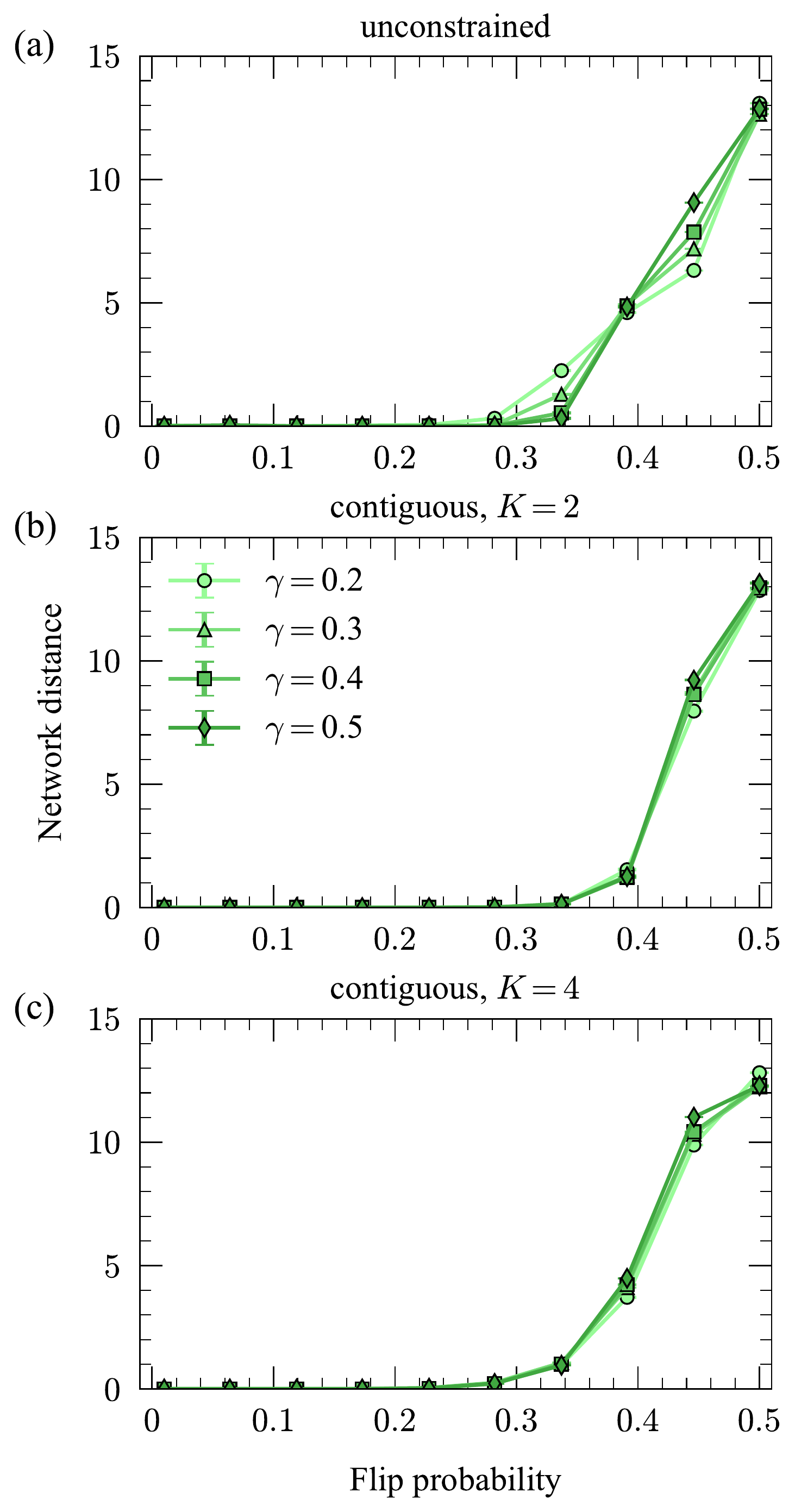}
    \caption{\textbf{Mode recovery for different mode separations}. Network distance between true and inferred modes for \textbf{(a)} unconstrained clustering, \textbf{(b)} contiguous clustering with $K=2$, and \textbf{(c)} contiguous clustering with $K=4$ for various values of the mode separation $\gamma$. Each data point is an average over $200$ realizations of the population for the corresponding value of the flip probability, and error bars correspond to three standard errors in the mean. }
    \label{fig:recovery2hamming}
\end{figure}

\section{Analysis of modes}
\label{sec:appendix_modes}

One of the advantages of using a concise set of modes instead of a complete collection of networks is that we can simplify analysis of the whole population greatly.
For example, a common task in network machine learning and statistics is network-level regression.
In such a task, our goal is to understand how the properties of networks relate to response variables, such as the adoption of a product or practice, or the efficiency of shipping.

To implement a regression, one denotes by $\bm{x}_i$ the vector of properties of network $i$ and by $y_i$ the response for this network, modeling:
\begin{equation}
    y_i = f(\bm{x}_i) + \epsilon_i
\end{equation}
where $\epsilon_i$ is a random variable modeling deviations from the predictor $f(\bm{x}_i)$.
Fitting the regression is a matter of identifying a good predictor function $f(\cdot)$ given data $\{(\bm{x}_1,y_1), (\bm{x}_2,y_2),...\}$

\begin{figure}
    \centering
    \includegraphics[width=0.9\linewidth]{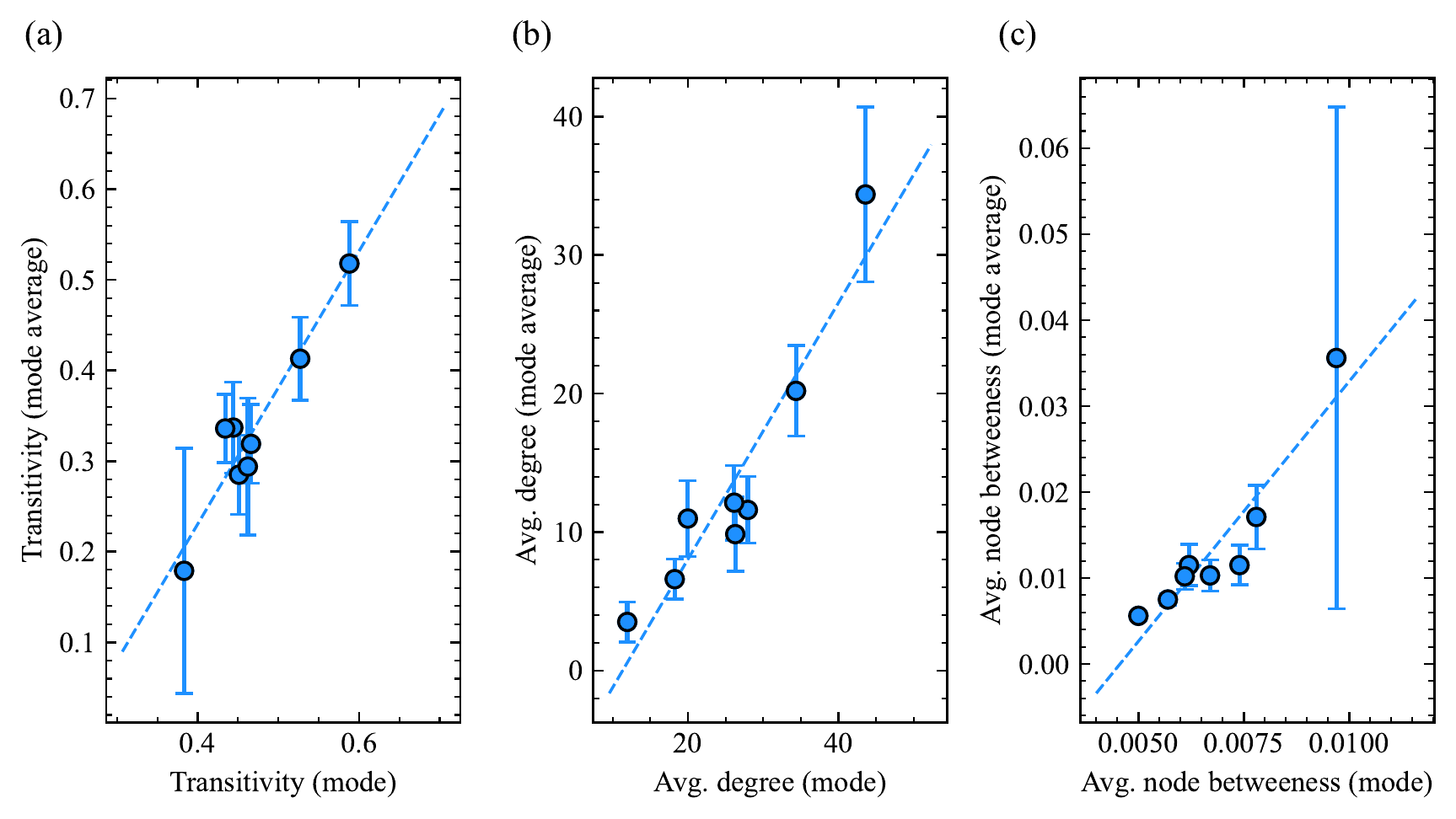}
    \caption{\textbf{Relationship between network predictors in modes and clusters of the FAO trade network.} Network quantities are computed for each of the 8 modes of the FAO trade network, and shown on the horizontal axis. The same quantities are then averaged for the networks of each associated cluster of networks, and shown on the vertical axis. The quantities are: (a) Transitivity, defined as a ratio of the number of triangles to the number of connected triplets. The Pearson correlation of the two quantities is 0.949. (b) Average degree, with a Pearson correlation coefficient of $0.943$. (c) Average node betweenness centrality, with a Pearson correlation coefficient of $0.936$. Error bars show one standard deviation of the quantities among the networks within each mode's cluster.}
    \label{fig:regression}
\end{figure}

Such analysis can become complex and impractical if the set of networks is too large (for example, if computing the network predictors is costly, as is the case of some centrality measures or macroscopic properties like community structure). 
Hence, one may run the analysis on a smaller set of representative samples.
Clustering with an information criterion can achieve this goal.
We demonstrate this in Fig.~\ref{fig:regression}, using the FAO Trade Network analyzed in Sec.~\ref{sec:food}.
The figure shows the relationship between the value of three representative network predictors as computed on the eight identified modes and the average value of these same predictors across the networks of the cluster associated with each mode.
One can see that the two quantities are proportional to one another: when, say, the average betweenness centrality of nodes is higher in the mode, it is also higher on average in the network of the associated cluster.
This means that we can run a regression analysis on a small set of modes instead of on a complete network---for example, by using the average response of the networks in any given node as the new response variable.

\section{Equivalence with Bayesian MAP estimation}
\label{sec:MAP}

In this appendix we establish the correspondence between our minimum description length objective and Bayesian MAP estimation in the heterogeneous network population model of \cite{YKN22Clustering}. Adapting to the notation in this paper and simplifying the product over network samples in Eq.~17 from \cite{YKN22Clustering}, the likelihood of the data $\mathcal{D}$ under the heterogeneous network population generative model is given by
\begin{align}
P(\mathcal{D}\vert \mathcal{A},\mathcal{C},\bm{\alpha},\bm{\beta})&=\prod_{k=1}^{K}\alpha_k^{t_k}(1-\alpha_k)^{S_kM_k-t_k}  \prod_{k=1}^{K}\beta_k^{p_k}(1-\beta_k)^{S_k M_k^\ast -p_k},    
\end{align}
where $\bm{\alpha}=\{\alpha_1,\alpha_2,...,\alpha_K\}$ and $\bm{\beta}=\{\beta_1,\beta_2,...,\beta_K\}$ are vectors of true- and false-positive rates within the clusters, and where we define $t_k = S_kM_k - n_k$ as the the number of true-positive edges, as in the previous Appendix.

We can additionally add priors on the modes and clusters to prepare for performing Bayesian inference with the model. As is done in \cite{YKN22Clustering}, for our prior on the clusters we can assume that each sample $s$ is assigned a cluster $C_k$ independently at random with prior probability $\pi_k$, giving the following prior over all clusters
\begin{align}
P(\mathcal{C}\vert \bm{\pi}) = \prod_{k=1}^{K}\pi_k^{S_k}. \end{align}
We can also suppose that each mode $\bm{A}^{(k)}$ has some prior probability $\rho_k$ for the existence of each edge $(i,j)$, which results in the following prior over all modes
\begin{align}
P(\mathcal{A}\vert \bm{\rho}) = \prod_{k=1}^{K}\rho_{k}^{M_k}(1-\rho_k)^{ M_k^\ast }.
\end{align}
This prior differs from the one used in \cite{YKN22Clustering}, which couples all modes together through a single edge probability $\rho$. This prior would result in a similar description length expression to ours when computing MAP estimators, except it does not have the beneficial properties of being a sum of decoupled cluster-level description lengths and allowing the modes to fluctuate independently in their densities. We finally assume (as in \cite{YKN22Clustering}) uniform priors over the parameters $\{\bm{\alpha},\bm{\beta},\bm{\pi},\bm{\rho}\}$, which we collectively denote as $\bm{\theta}$. 

Applying Bayes' rule, the posterior distribution over modes $\mathcal{A}$, clusters $\mathcal{C}$, and model parameters $\bm{\theta}$ is given by
\begin{align}
\label{eq:posterior}
P(\mathcal{A},\mathcal{C},\bm{\theta}\vert \mathcal{D}) &=
\frac{P(\mathcal{D}, \mathcal{A},\mathcal{C},\bm{\theta})}{\sum_{\mathcal{A},\mathcal{C},\bm{\theta}}P(\mathcal{D}, \mathcal{A},\mathcal{C},\bm{\theta})} \nonumber\\
&\propto P(\mathcal{D}\vert \mathcal{A},\mathcal{C},\bm{\alpha},\bm{\beta})P(\mathcal{C}\vert \bm{\pi})P(\mathcal{A}\vert \bm{\rho}).   \end{align}
Now, to find the Maximum A Posteriori (MAP) estimators $\{\mathcal{\hat A},\mathcal{ \hat C},\bm{\hat\theta}\}$, we can maximize Eq.~\eqref{eq:posterior} (or equivalently, its logarithm) using the method of profiling. To do this, we first maximize over the parameters $\bm{\theta}$, finding the MAP estimates $\bm{\hat\theta}(\mathcal{A},\mathcal{C})$ as a function of the configuration $\{\mathcal{A},\mathcal{C}\}$. Then, we can plug the function(s) $\bm{\hat\theta}(\mathcal{A},\mathcal{C})$ back into Eq.~\eqref{eq:posterior} to obtain an objective function over the configuration $\{\mathcal{A},\mathcal{C}\}$, which can then be optimized over the remaining variables to find all the MAP estimators.

The logarithm of the posterior distribution in Eq.~\eqref{eq:posterior}, which can be maximized to find the MAP estimators, is given by
\begin{align}
\label{eq:logposterior}   
\log P(\mathcal{A},\mathcal{C},\bm{\theta}\vert \mathcal{D}) &= \sum_{k=1}^{K}\left[t_k\log \alpha_k + (S_kM_k-t_k)\log (1-\alpha_k)\right] \nonumber\\
&+ \sum_{k=1}^{K}\left[p_k\log \beta_k + (S_k M_k^\ast -p_k)\log (1-\beta_k)\right]\nonumber\\
&+\sum_{k=1}^{K}S_k\log \pi_k\\
&+\sum_{k=1}^{K}\left[M_k \log \rho_k +  M_k^\ast  \log (1-\rho_k)\right]. \nonumber
\end{align}
Maximizing over the model parameters $\bm{\theta}$ gives the estimators
\begin{align}
\hat \alpha_k &= \frac{t_k}{S_kM_k}  \\
\hat \beta_k &= \frac{p_k}{S_k M_k^\ast }\\
\hat \pi_k &= \frac{S_k}{S}  \\
\hat \rho_k &= \frac{M_k}{{N\choose 2}}  
\end{align}
These can be derived using the method of Lagrange multipliers, with the normalization constraint $\sum_{k=1}^{K}\pi_k=1$ for the group assignment probabilities. Finally, replacing these expressions back into Eq.~\eqref{eq:logposterior}, and using $t_k= S_kM_k-n_k$ we find
\begin{align}
\label{eq:Mapequiv}
\log P(\mathcal{A},\mathcal{C},\bm{\hat\theta}(\mathcal{A},\mathcal{C})\vert \mathcal{D}) 
= -\mathcal{L}(\mathcal{D}).
\end{align}
Thus, maximizing Eq.~\eqref{eq:Mapequiv} to find (profiled) MAP estimators of this Bayesian model is equivalent to minimizing the description length $\mathcal{L}(\mathcal{D})$. 
Further, with $\bm{\theta}=\hat{\bm{\theta}}$, the first two lines of Eq.~\eqref{eq:logposterior} correspond to $\mathcal{L}(\mathcal{D}|\mathcal{A},\mathcal{C})$, the third line corresponds to $\mathcal{L}(C)$, and the last line to $\mathcal{L}(A)$, which establishes a direct mapping between likelihoods, priors and each part of the multi-part encoding.
This correspondence unifies the Bayesian and MDL formulations of network population clustering and explains the good reconstruction performance seen in the experiments of the main text~\cite{mackay2003information,grunwald2007minimum}.

\section{Additional related methods for clustering network populations}
\label{sec:comparison}

The two algorithms proposed in this paper are not the first methods capable of clustering network populations. Indeed, as discussed in the Introduction, there are a variety of existing methods that have tackled this problem using various modeling assumptions. However, directly making an apples-to-apples numerical comparison of the performance of these methods with our own is difficult for a number of reasons. Firstly, as mentioned in the Introduction, most of these methods do not have a unified model selection criterion that allows for automatic inference of the number of clusters. Therefore to make a completely fair numerical comparison with our own methods, we would either have to fix the number of clusters in all methods being compared---removing the additional obstacle of model selection which is one of the key innovations of our method---or add an ad hoc regularization term/Bayesian prior to the existing methods, to which their output may be highly sensitive \cite{mantziou2021}. Additionally, class labels of real-world network datasets are often arbitrary and uncorrelated with the structure of the data~\cite{peel2017ground}, so any purely numerical comparison of unsupervised clustering algorithms for network data using label recovery in real-world examples will be ill-posed \cite{peixoto2022implicit}. Recovery of synthetic clusters, as described in the experiment in Sec.~\ref{sec:synthetic}, can serve as a well-posed task for testing the relative performance of existing algorithms if the data can be generated from the model corresponding to each algorithm \cite{peixoto2022implicit}, but many existing methods do not have explicit generative models and fitting them to other synthetic data may result in severe model misspecification \cite{spanos1986statistical}. 

\begin{figure}
    \centering
    \includegraphics[width=0.5\linewidth]{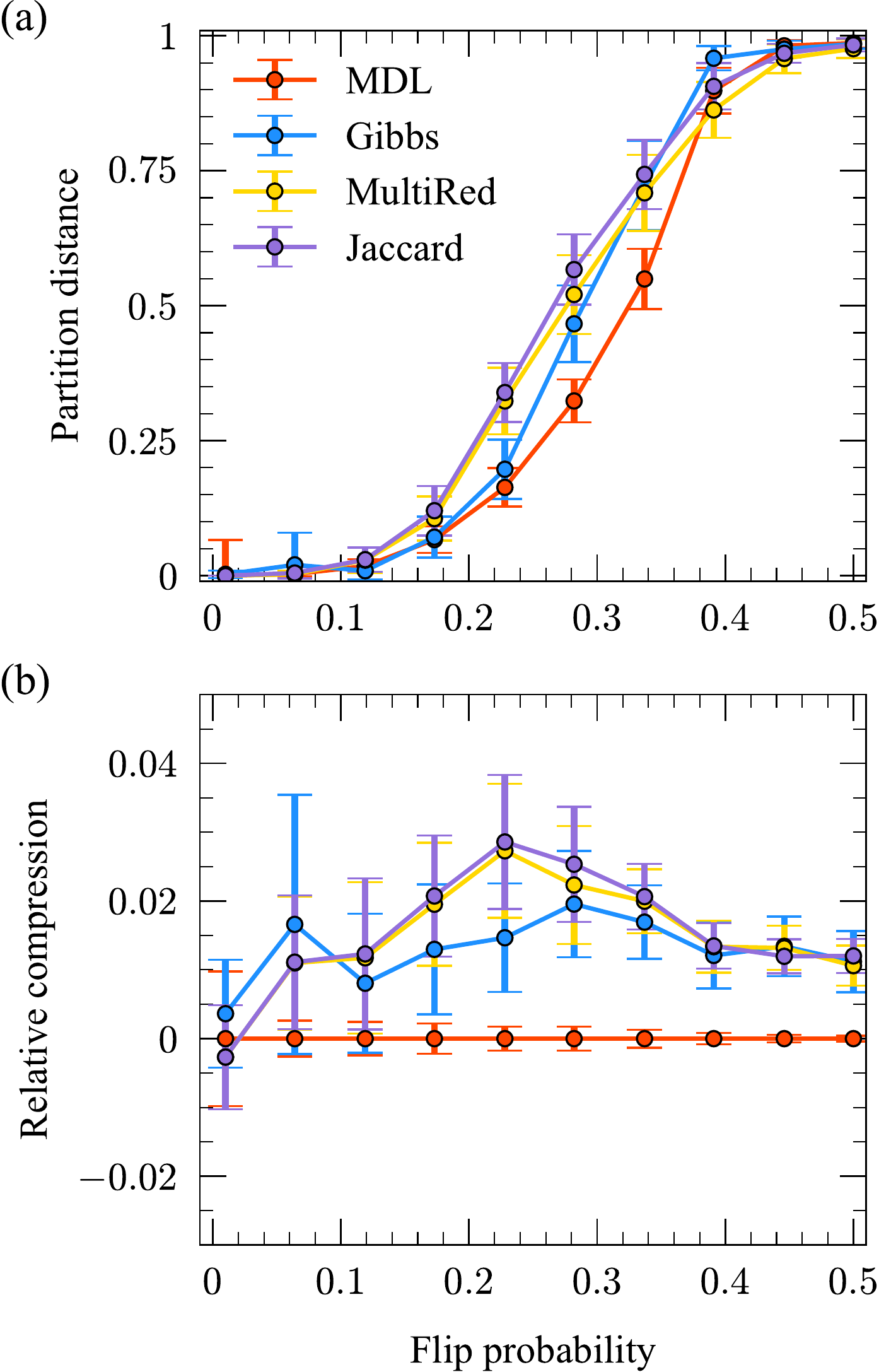}\\
    \caption{\textbf{Comparison of reconstruction performance across alternative layer clustering approaches.} \textbf{(a)} Partition distance and \textbf{(b)} difference in mean inverse compression ratio (Eq.~\eqref{eq:ratio}) with the method of this paper. Each data point is an average over $50$ realizations of the population for the corresponding value of the flip probability. Error bars correspond to three standard errors in the mean. } 
    \label{fig:algcomparison}
\end{figure}

With these caveats in mind, we can assess how our proposed algorithm is performing in comparison with similar existing methods by applying other algorithms to the same synthetic discontiguous reconstruction task as in Fig.~\ref{fig:recovery}. (No directly comparable methods exist for the temporally contiguous compression task.) Since alternative layer/population clustering algorithms are also seeking similar structural regularities in the data, we would expect these to perform well reconstructing the same synthetic benchmark used to verify the performance of the proposed method. In Fig.~\ref{fig:algcomparison}, we show the results of these experiments, comparing the reconstruction performance of our method with the methods of~\cite{YKN22Clustering} (denoted ``Gibbs''),~\cite{de2015structural} (denoted ``MultiRed''), and a naive baseline (denoted ``Jaccard'') where layers are clustered by applying the Ward hierarchical linkage criterion (used in~\cite{de2015structural}) to the pairwise matrix of Jaccard distances among the layers' edge sets. These methods are chosen for comparison as they are the most similar to the proposed method---the method of~\cite{YKN22Clustering} tackles a similar inference problem but from a Bayesian perspective and with a different prior on the mode densities; the method of~\cite{de2015structural} aims to compress the information in the network layers but uses the Jensen-Shannon Divergence between network Laplacians; and the naive baseline also uses the overlap of the edge positions among networks to identify structurally cohesive clusters. 

None of these methods constructs a single unweighted representative modal network for each cluster as our method does, so to compute the description lengths for each method, we apply the greedy mode update algorithm of Appendix~\ref{sec:appendixmergesplit} to the identified clusters of network layers and evaluate the resulting configuration of modes and clusters using Eq.~\ref{eq:DLtotal}. However, since no principled means exists to construct comparable modal networks of alternative methods without imposing these assumptions, we do not compare the mode reconstruction performance in this experiment (panel (a) in Fig.~\ref{fig:recovery}). We have also given a significant advantage to the alternative algorithms---we have fixed their inferred number of clusters to the correct value $K=2$. In contrast, our algorithm must learn the value of $K$ automatically. All algorithms perform well on the synthetic benchmark in identifying the planted clusters, but our algorithm performs substantially better for intermediate flip probabilities despite needing to identify the correct value of $K$ (Fig.~\ref{fig:algcomparison}a). Since our method aims to minimize the description length used to evaluate the algorithms, our method has an edge in terms of compression  (Fig.~\ref{fig:algcomparison}b).   

Below we provide a further qualitative comparison between our methods and existing methods to demonstrate the substantial innovations and benefits of our proposed algorithms: 
\begin{itemize}
    
    \item A number of recent methods propose to cluster populations of networks by imposing structure on the clusters with explicit network models such as stochastic block models, exponential random graphs, or latent space models~\cite{stanley2016clustering,mantziou2021,yin2022finite,durante2017nonparametric}. This approach allows the user to simultaneously infer the clusters of networks as well as the model parameters for each cluster. While the model parameters within each cluster can be useful for identifying common structure among its constituent networks, any explicit model will necessarily miss out on important structure (in particular, transitivity is often not present in parametric network models). Explicit model inference also often necessitates a complex and time intensive within-cluster estimation scheme such as expectation maximization to identify stochastic block model communities and model parameters at the network-cluster level. In contrast, our method takes a non-parametric approach for the representative cluster summaries and infers a single representative modal network for each cluster using a simple greedy edge removal approach, which is only constrained in its density since this will affect modal information content.

    \item Other methods perform network change point detection (\cite{peel2015detecting} and \cite{peixoto2018change}), which can be seen as equivalent only to the objective of contiguous partitioning we approach in this paper. On the other hand, the information theoretic methods of this paper unify the discontiguous and contiguous clustering objectives, only modifying a single term of the encoding to accommodate the more restrictive setting of the latter clustering problem. The contiguous clustering method of this paper also has the benefit of achieving exact optimization (up to the greedy mode updates) in polynomial time. Change point detection methods also may allow for significant but slow temporal variation between change points, resulting in very heterogeneous clusters, since they only look for substantial jumps in network structure across time.

    \item The methods of \cite{nielsen2018multiple}, \cite{wang2019joint}, \cite{arroyo2021inference} use joint embeddings of network populations. These methods use a single vector to model each node's position in a latent space across all networks in the population, while the loadings of the latent space decomposition may change between clusters to capture heteorgeneity across network layers. These methods have similar advantages and disadvantages as the methods \cite{stanley2016clustering,mantziou2021,yin2022finite,durante2017nonparametric} discussed above when compared to our methods. 
    
    \item In \cite{la2016gibbs}, a Gibbs distribution approach is used to model network populations, where the Hamiltonian determining the properties of the ensemble is based on a user-specified graph distance to the modal network. This approach, like our own, infers an entire graph for each mode, giving it the flexibility to capture a range of possible network structures. On the other hand, this method requires the user to specify a graph distance of interest, of which there are many to choose from and to which the results are sensitive, in contrast with our method which requires no user input. The same downside arises when using the Fr\'echet mean \cite{lunagomez2020modeling} to define a modal network of a cluster.
    
    \item The method of \cite{YKN22Clustering} uses Gibbs sampling to construct posterior distributions over modal edge probabilities and cluster assignments. As with some other methods discussed, this method requires the user to specify an ad hoc prior on the modal structures and re-run the time-consuming sampling process for multiple values of $K$ to infer the optimal number of clusters. The particular prior used in \cite{YKN22Clustering} also has the disadvantage of coupling all the modes together with respect to their densities, whereas our transmission scheme treats all clusters independently.

    \item The method in \cite{de2015structural} reduces multilayer networks into a smaller set of aggregated layers using the network Jensen-Shannon Divergence as a measure of information redundancy. While somewhat conceptually similar to our own approach in that it is based in information theory, this method does not explicitly construct parsimonious modal structures describing the structure of the clusters, and instead sums all layers within a cluster. It also performs pairwise comparisons using the Jensen-Shannon Divergence---rather than group-wise comparisons as done, for example, in \cite{kirkley2020information}---which may result in biased estimation of information redundancies in comparison to our method. This could potentially explain the method returning many more clusters than would be optimal for our method (see Sec.~\ref{sec:food}). Finally, this method requires computing the layers' Laplacian eigenspectra, which results in a significant increase in computation time compared to our method, even when approximating the spectra with polynomial approximations.

    \item The approach used in \cite{rojas_multiscale_2021}, like some other multilayer community detections methods, clusters a multilayer network into communities of nodes within and across layers, so it is not directly comparable to our own method. It does however provide a complementary perspective to the analysis in Sec.~\ref{sec:fossils}.

\end{itemize}

While these algorithms we discuss all demonstrate good performance for certain aspects of multilayer network/network population clustering, the methods of this paper have two significant advantages over all algorithms discussed above. The first is that these methods do not have a natural criterion for specifying the number of clusters $K$. Therefore, model selection needs to be addressed with ad hoc or approximative regularization methods (e.g., arbitrary information criteria, sparse finite mixtures) instead of the principled automatic penalization of extra modes we achieve through a minimum description length formulation. As discussed in~\cite{mantziou2021}, the choice of penalty for the number of clusters significantly impacts the results, so this model selection is a critical innovation of our method. Second, our proposed dynamic programming scheme for solving the restricted problem of contiguous layer clustering is unique among existing methods, which do not consider the restriction to contiguous clusters.

\end{document}